\documentclass[journal=jpclcd,manuscript=letter]{achemso}

\usepackage[version=3]{mhchem} 
\usepackage[table]{xcolor}
\usepackage[version=3]{mhchem} 
\usepackage{graphicx} 
\usepackage{subcaption} 
\usepackage{hyperref}
\usepackage{color}
\usepackage{amssymb}
\usepackage{amsmath}

\mathchardef\mhyphen="2D

\author{Rugwed A. Lokhande}
\affiliation{Department of Chemistry, University of Florida, Gainesville, FL 32603, USA}
\alsoaffiliation{Quantum Theory Project, University of Florida, Gainesville, FL 32603, USA}
\author{Carlos E. V. de Moura}
\affiliation{Department of Chemistry, University of Florida, Gainesville, FL 32603, USA}
\alsoaffiliation{Quantum Theory Project, University of Florida, Gainesville, FL 32603, USA}
\author{Pratiksha B. Gaikwad}
\affiliation{Department of Chemistry, University of Florida, Gainesville, FL 32603, USA}
\alsoaffiliation{Quantum Theory Project, University of Florida, Gainesville, FL 32603, USA}
\author{Ramón Alain Miranda-Quintana}
\affiliation{Department of Chemistry, University of Florida, Gainesville, FL 32603, USA}
\alsoaffiliation{Quantum Theory Project, University of Florida, Gainesville, FL 32603, USA}
\email{quintana@chem.ufl.edu}

\email{quintana@chem.ufl.edu}

\title{Exploring new construction schemes for extended-hierarchy CI (ehCI) wavefunctions}

\abbreviations{}

\begin{document}







\begin{abstract}
Following the work of Loos et al. on hierarchy CI (hCI) we propose an extended way of partitioning the Hilbert space by combining the excitation and the seniority sectors in a more general way. We define the hierarchy parameter, h, involving two ‘weights’ (alphas), measuring the importance of excitation ($e$) and seniority ($s$) contributions to the wavefunction according to: $h= \alpha_1.e +\alpha_2.s$. We test these new CI methods on the dissociation of chemical systems and analyze them based on the importance of these contributions. 
\end{abstract}

\section{Introduction}

Accurately describing electron wavefunctions is among the most persistent challenges in \emph{ab initio} quantum chemistry.
A formally exact solution is obtained by diagonalizing the electronic Hamiltonian in the full Fock‐space spanned by all Slater determinants that can be built from a chosen one–particle basis.\cite{sa1996,ht2013}
This full configuration–interaction (FCI) limit, however, is only accessible for the smallest of the molecular systems involving 20 or fewer electrons because the size of the Hilbert space grows combinatorially with the number of spin‐orbitals.\cite{kp1984, kp1989}
Since a full diagonalization of the entire determinant space quickly becomes computationally intractable, we need systematic ways to select a much smaller set of electron arrangements (Slater determinants) that still capture most of the correlation energy, i.e. the difference between the exact energy and the simpler Hartree–Fock (HF) result.

A common strategy for selecting important Slater determinants is to organize them based on the number of electrons that are excited relative to a chosen reference determinant, which is typically the Hartree-Fock determinant, though this is not strictly required.
This gives rise to a hierarchy of truncated Configuration Interaction (CI) methods: including all single excitations yields CIS, adding double excitations leads to CISD, and so on (e.g., CISDT, CISDTQ).\cite{pople1976theoretical,br2009}
While the number of determinants in these models increases polynomially with the size of the basis set for each excitation rank, the prefactor becomes prohibitively large at higher truncation levels, rendering even CISD impractical for systems of moderate size.
Excitation‐based CI accurately recovers dynamic correlation(short-range, instantaneous electron repulsion). However, these models are not size-consistent and struggle when the system becomes strongly correlated, for e.g. when the chemical bonds stretch or when several electronic states lie close in energy. \cite{bjs1978, brj1982}

A different idea is to group determinants by their seniority ($\Omega$), which is the number of orbitals that contain a single (unpaired) electron.
Restricting the wavefunction to the seniority zero sector, in which each spatial orbital is doubly occupied, we obtain doubly occupied CI (DOCI).
Many studies have explored DOCI and shown that it reliably captures static (strong) correlation. \cite{sg2011,sg2014, sg_2014_1, bp2013,sg2015}
DOCI is a size-consistent method that accurately captures static (non-dynamical) correlation effects, which are particularly pronounced in situations such as bond dissociation.
Despite its advantages, DOCI exhibits exponential scaling in the number of determinants with system size.
Moreover, it does not account for a substantial portion of the dynamical correlation energy when used in isolation. \cite{bp2015}

Coupled-cluster (CC) theory achieves size-extensivity by expressing the wavefunction as an exponential of excitation operators acting on a reference determinant, thereby systematically incorporating electron correlation based on excitation ranks.
The standard coupled-cluster approach, including singles and doubles with a perturbative treatment of triples (CCSD(T)), typically achieves near-chemical accuracy for systems dominated by dynamical correlation. \cite{br2007, shf2000}
However, its accuracy deteriorates significantly in the presence of strong static correlation or near-degeneracies, where the single-reference assumption breaks down and the perturbative triples correction becomes invalid.\cite{br2007,hgm1989}  

In recent years, selected CI methods such as Configuration Interaction using a Perturbative Selection made Iteratively (CIPSI) and heat bath CI have become popular.
CIPSI constructs the wavefunction iteratively by selecting determinants with the largest estimated contributions to the correlation energy, often achieving near-FCI accuracy with a substantially reduced configuration space.\cite{ucj2016,wkb2016}
However, these methods involve diagonalization procedures of large Hamiltonian subspaces, making them computationally demanding for extended reaction pathways or systems with large active spaces.

Although seniority-based CI methods exhibit better performance in capturing static correlation compared to excitation-based CI methods, they come with a higher computational cost. \cite{rk2015, sge2008_1}
In contrast, excitation-based methods perform better in capturing dynamic correlation efficiently and at a polynomial cost relative to the system size.
Our objective is to develop a method capable of effectively capturing both static and dynamic correlations with minimal determinants.
Loos et al. earlier work introduced \emph{hierarchy configuration–interaction} (hCI) as a way to combine the strengths of excitation-based and seniority-based pictures.\cite{lpf2022}
Each determinant receives a score
\begin{equation}
    \textit{h}= \dfrac{\textit{e} + \textit{s}/2}{2}
\end{equation}
where $e$ is the excitation rank and $s$ is the seniority.
All determinants with $h\leq h_{\max}$ form a nested and variational space that can be systematically converged toward FCI.

We introduce a more generalized approach called extended hierarchy CI (ehCI), which combines the excitation degree (\textit{e}) and seniority number (\textit{s}) in a weighted manner into a single hierarchy parameter (\textit{h}). 
\begin{equation}
    \textit{h}={\alpha_{1}\textit{e} + \alpha_{2}\textit{s}}
\end{equation}

\begin{figure}[H] 
    \centering
    \includegraphics[scale=0.8]{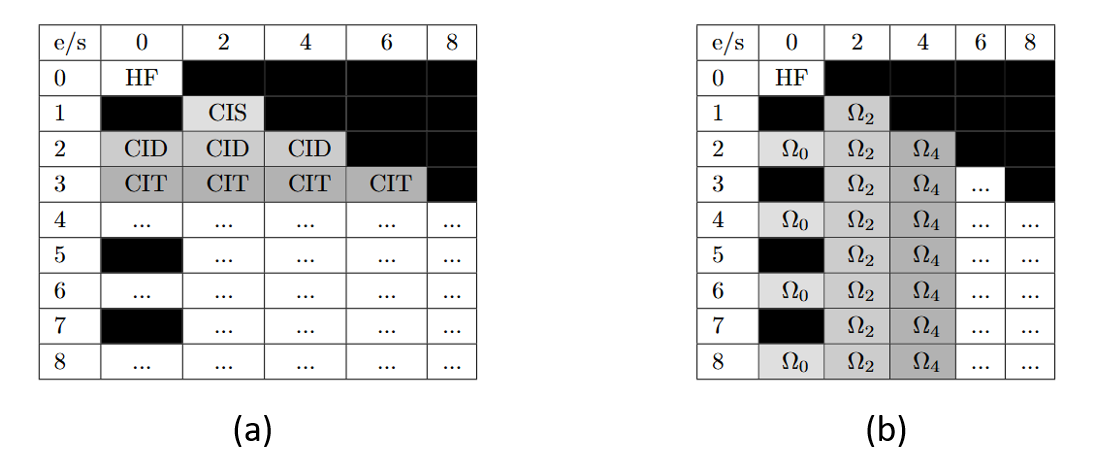} 
    
    \caption{This figure illustrates the case of an eight-electron system initialized from a closed-shell reference determinant. The Hilbert space divided into blocks based on two properties: the excitation degree \textit{e} (relative to a closed-shell reference determinant) and the seniority number \textit{s}.
    (a) Standard excitation-level CI methods (e.g., CISD, CISDT) correspond to parameters
$\alpha_1=1$ and $\alpha_2=0$, which map onto the rows of the diagram.
    (b) Seniority-based CI corresponds to $\alpha_1=0$ and $\alpha_2=1$, the columns of the diagram.}
\label{fig:ehCI}
\end{figure}

\begin{figure}[H] 
    \centering
    \includegraphics[scale=0.75]{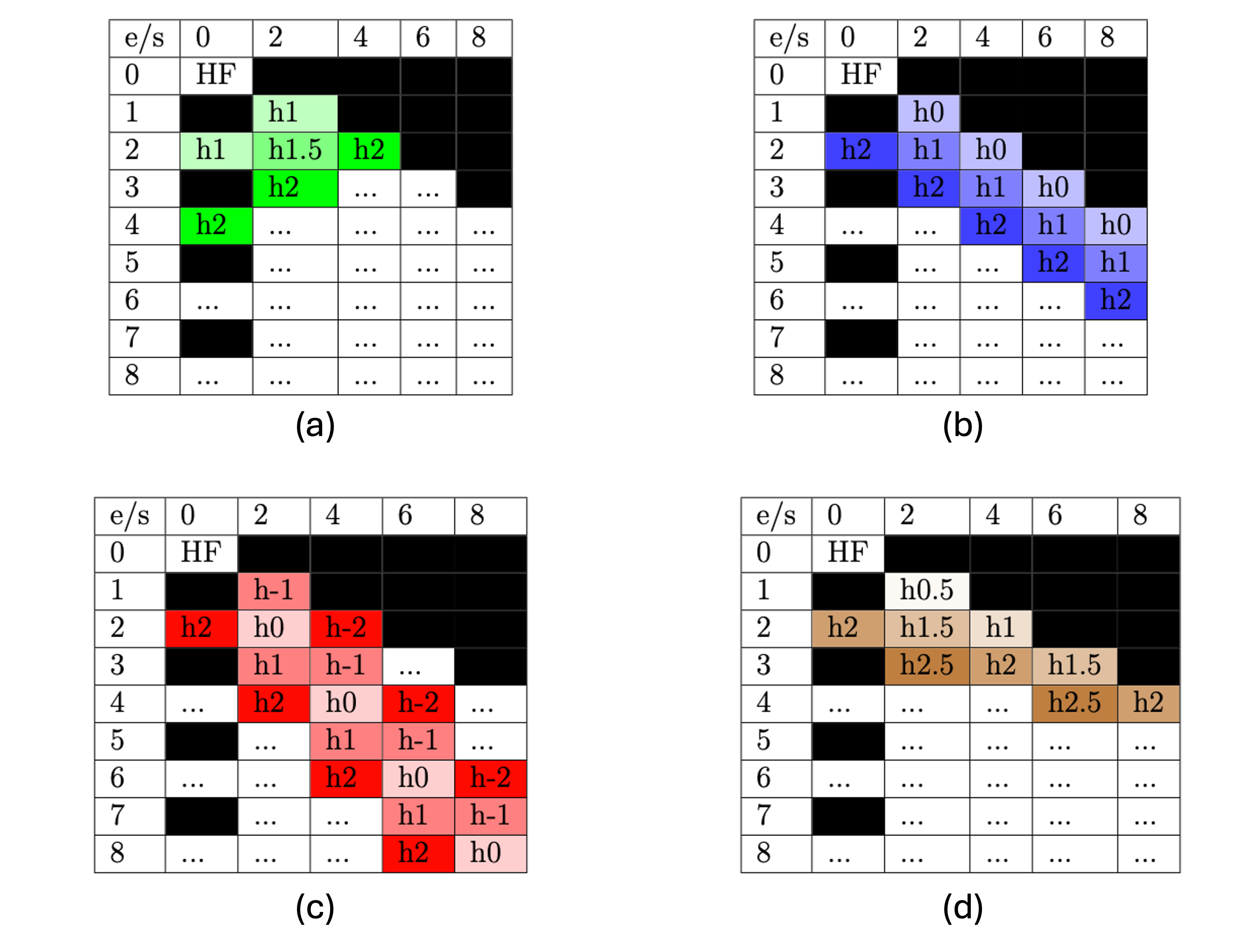} 
    
    \caption{Four different partitioning schemes of the Hilbert spaces. 
    (a) Positive slope diagonals (PSD) (b) Negative slope diagonals (NSD) (c) Vertical chess horse (VCH) (d) Horizontal chess horse (HCH)}
\label{fig:patterns}
\end{figure}

\autoref{fig:ehCI} illustrates the FCI space for an eight-electron system referenced to a closed-shell determinant, with the Hilbert space divided into blocks characterized by excitation degree \(e\) and seniority \(s\).
Inaccessible configurations are shown as black squares, while accessible determinants (denoted by “\dots”) span the FCI manifold. 
Beyond the two limiting truncations, the standard excitation-level CI (CISD, CISDT, \(\alpha_1=1,\alpha_2=0\), corresponding to horizontal rows in \autoref{fig:ehCI}) and the seniority-based CI (\(\alpha_1=0,\alpha_2=1\), corresponding to vertical columns), we consider several mixed partitions of the Hilbert space, each defined by a linear hierarchy parameter.
\autoref{fig:patterns} presents several such configurations, including ``Positive Slope Diagonals'' (PSD),  ``Negative Slope Diagonals'' (NSD), ``Vertical Chess Horse'' (VCH), and ``Horizontal Chess Horse''(HCH). The ``Horse'' nomenclature is based upon how a horse would move on a chessboard.
Of these, the Positive Slope Diagonals (PSD) pattern was originally introduced by Loos.
This PSD weighting naturally balances excitation degree and seniority and motivates the exploration of alternative \((\alpha_{1},\alpha_{2})\) combinations to optimize both dynamic and static correlation recovery.


To construct the hierarchy corresponding to the ``Negative Slope Diagonals'' shown in \autoref{fig:patterns}, it is necessary to identify the appropriate weighting parameters \(\alpha_1\) and \(\alpha_2\) that define the desired partitioning of configuration space.
Consider two representative configurations lying along one of the longest such diagonals (highlighted in light blue), specifically \((e = 1, s = 2)\) and \((e = 2, s = 4)\).
For these blocks to belong to the same hierarchy level \(h\), the following condition must be satisfied:
\begin{equation}
    \alpha_1 \cdot 1 + \alpha_2 \cdot 2 = \alpha_1 \cdot 2 + \alpha_2 \cdot 4,
\end{equation}
which simplifies to:
\begin{equation}
    \alpha_2 = -0.5 \cdot \alpha_1.
\end{equation}

Since the absolute magnitudes of \(\alpha_1\) and \(\alpha_2\) are unimportant for defining the hierarchy, they can be normalized within an arbitrary interval (e.g., \([-1, 1]\)). Choosing \(\alpha_1 = 1\) and \(\alpha_2 = -0.5\) leads to a hierarchy function of the form:
\begin{equation}
    h = e - 0.5s.
\end{equation}

Similar procedures can be applied to derive the \(\alpha\) parameters for other partitioning schemes.
A summary of the resulting \(\alpha_1\), \(\alpha_2\), and corresponding hierarchy equations for various partitions is provided in \autoref{tab:ehCI_table}.

\begin{table}[h!]
\centering
\begin{tabular}{|c|c|c|c|c|}
\hline
\textbf{No.}&\textbf{Partition} & \textbf{$\alpha_1$} & \textbf{$\alpha_2$} & \textbf{Hierarchy equation}\\ \hline
1    & Positive Slope Diagonals (PSD)      & 0.5     & 0.25    & $h = 0.5e + 0.25s$ \\ \hline
2     & Negative Slope Diagonals (NSD)   & 1     & -0.5    &$h = e - 0.5s$\\ \hline
3     & Vertical Chess Horse (VCH)    & 1    & -1    & $h = e - s$\\ \hline
4     & Horizantal Chess Horse (HCH)    & 1      & -0.25 &   $h = e - 0.25s $\\ \hline
\end{tabular}
\caption{Different partitioning of the Hilbert space with corresponding parameters alpha and the hierarchy equation}
\label{tab:ehCI_table}
\end{table}

\section{2. MODEL SYSTEMS AND COMPUTATIONAL DETAILS}
\begin{figure}[H] 
    \centering
    \includegraphics[scale=0.45]{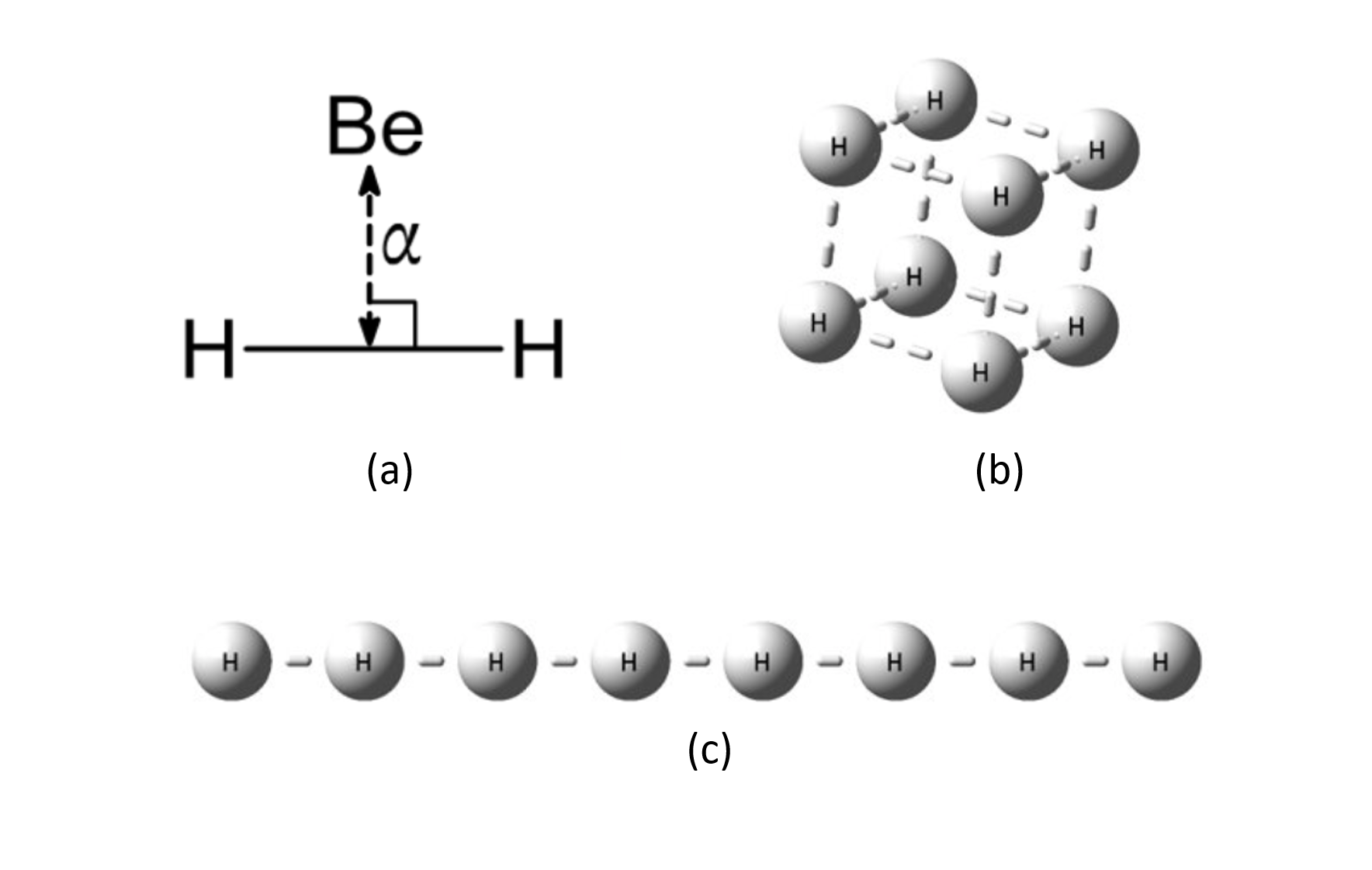} 
    
    \caption{Model systems: (a) \ce{BeH2}, (b) cubic \ce{H8} and (c) linear \ce{H8}.}
\label{fig:image1}
\end{figure}

To benchmark the performance of the newly developed ehCI method, we evaluated it on several well-established model systems, including the classic $C_{2v}$ insertion of \ce{Be} atom into \ce{H2} molecule, and the eight-electron hydrogen systems (\autoref{fig:image1}).\cite{pg1983}
The \ce{BeH2} molecule, with its closely spaced 2s and 2p orbitals (a phenomenon known as quasi-degeneracy), exhibits a combination of weak and strong electron correlation effects.
To explore these characteristics, we analyzed ten distinct molecular geometries inspired by a prior study by Purvis et al.\cite{pg1983}
Each geometry was constructed by varying two key structural parameters: the \ce{H}–\ce{H} bond length and the perpendicular distance of the Be atom from the \ce{H2} plane.
In these configurations, the \ce{Be} atom was placed at the origin of the coordinate system, and the Cartesian coordinates of the hydrogen atoms for each geometry are summarized in \autoref{tab:coordinates}.
The energy dependence on the perpendicular distance between \ce{Be} and \ce{H2} is illustrated in the resulting plots.
\begin{table}[h!]
\centering
\begin{tabular}{|c|c|c|c|}
\hline
\textbf{Point} & \textbf{X}   & \textbf{Y}          & \textbf{Z}   \\ \hline
A              & 0.0          & $\pm$2.54           & 0.0          \\ \hline
B              & 0.0          & $\pm$2.08           & 1.0          \\ \hline
C              & 0.0          & $\pm$1.62           & 2.0          \\ \hline
D              & 0.0          & $\pm$1.39           & 2.5          \\ \hline
E              & 0.0          & $\pm$1.275          & 2.75         \\ \hline
F              & 0.0          & $\pm$1.16           & 3.0          \\ \hline
G              & 0.0          & $\pm$0.93           & 3.5          \\ \hline
H              & 0.0          & $\pm$0.70           & 4.0          \\ \hline
I              & 0.0          & $\pm$0.70           & 6.0          \\ \hline
J              & 0.0          & $\pm$0.70           & 20.0         \\ \hline
\end{tabular}
\caption{Cartesian coordinates of \ce{BeH2} geometry for different points in Bohr units.}
\label{tab:coordinates}
\end{table}

For the eight-electron systems, two distinct models were investigated using eight hydrogen atoms: a linear chain arrangement and a cubic configuration.
In the linear \ce{H8} chain investigation, different geometries were generated by systematically varying the internuclear distances between adjacent hydrogen atoms in the range of [0.5, 4] Angstrom.
For the cubic \ce{H8} system, the cube edges were displaced from [0.5, 4.0] Bohr in length.
Structural details for this model system were adapted from Ref. \cite{grc2016}

All computational methods described in this study were implemented using \textsc{Fanpy} \cite{ap2023, fanpy_github}, an open-source Python 3 package designed to facilitate the development, testing, and application of advanced electronic structure methods within the \textsc{FANCI} framework \cite{ap2021, ap2024, mqr2024}.
\textsc{Fanpy} provides a flexible and modular infrastructure for prototyping novel wavefunction ansätze and automating the evaluation of their corresponding energy expressions, making it particularly well-suited for exploring new ab initio methodologies.
The calculations were done using the STO-6G basis set,\cite{pja_1969} with Molecular Orbitals (MOs) generated from Restricted Hartree–Fock (RHF) calculations.
The one- and two-body Hamiltonian integrals were processed using the HORTON library.
\cite{ap2024_1}
For the \ce{H8} clusters and the \ce{BeH2} model system, Full Configuration Interaction (FCI) energies were computed using \textsc{Fanpy} to serve as reference benchmarks for evaluating the accuracy of the energies obtained from the ehCI method.

\section{Results and Discussion}
\subsection{\ce{BeH2} curves}
In this section, we analyze the ground‐state potential energy curve (PEC) and associated errors for the C$_{2v}$ insertion of \ce{Be} into \ce{H2}, linear \ce{H8} dissociation, and cubic \ce{H8} dissociation, computed using the four newly defined ehCI methods (\autoref{tab:ehCI_table}).
We compare each ehCI variant to FCI, all calculated using the STO-6G basis.
Each scheme is defined by a hierarchy parameter 
\[
h \;=\; \alpha_{1}e \;+\; \alpha_{2}s,
\]
where $e$ is the excitation level (number of electrons excited) and $s$ is the seniority (number of unpaired electrons) of a determinant.
For a given maximum hierarchy cutoff $h_{\mathrm{max}}$, each partition scheme includes all determinants whose $h$ value does not exceed $\lvert h_{\mathrm{max}} \rvert$.
Because the coefficients $(\alpha_{1},\alpha_{2})$ differ among the four partition schemes, the number and type of determinants included at each hierarchy number vary. For instance, when \(\alpha_{2}<0\) (in NSD), the hierarchy parameter
\[
h \;=\;\alpha_{1}e + \alpha_{2}s
\]
prioritizes the inclusion of higher‐seniority determinants at lower values of \(h\), with lower‐seniority configurations entering only as \(h\) increases. Conversely, when \(\alpha_{2}>0\) (in PSD), lower‐seniority determinants are admitted at smaller \(h\), deferring the addition of higher‐seniority configurations until \(h\) exceeds the appropriate threshold.
\begin{figure}[H] 
    \centering
    \begin{subfigure}[b]{0.45\textwidth} 
        \centering
        \includegraphics[scale=0.335]{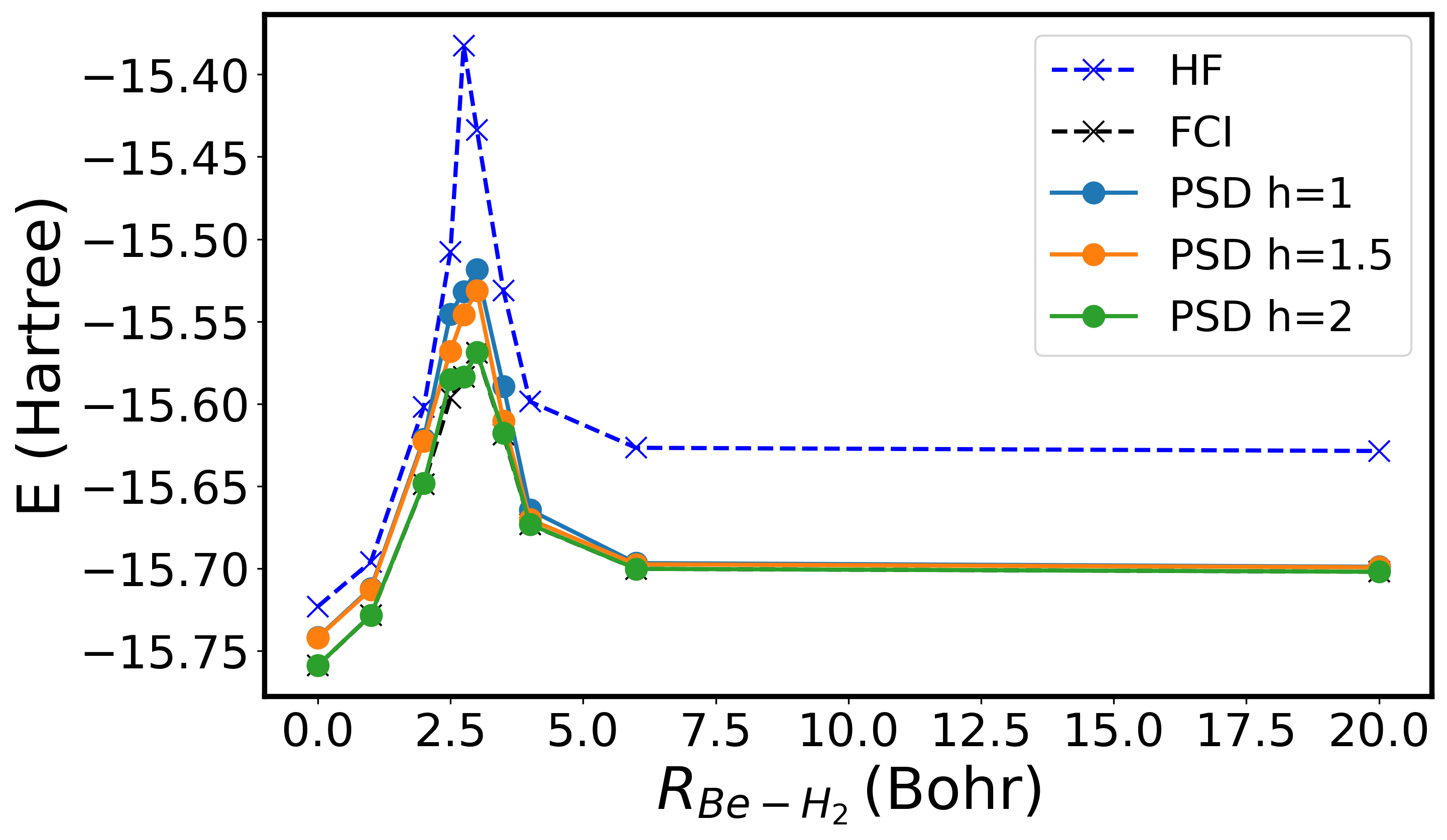} 
        \label{fig:image2}
    \end{subfigure}
    \hfill
    \begin{subfigure}[b]{0.45\textwidth} 
        \centering
        \includegraphics[scale=0.335]{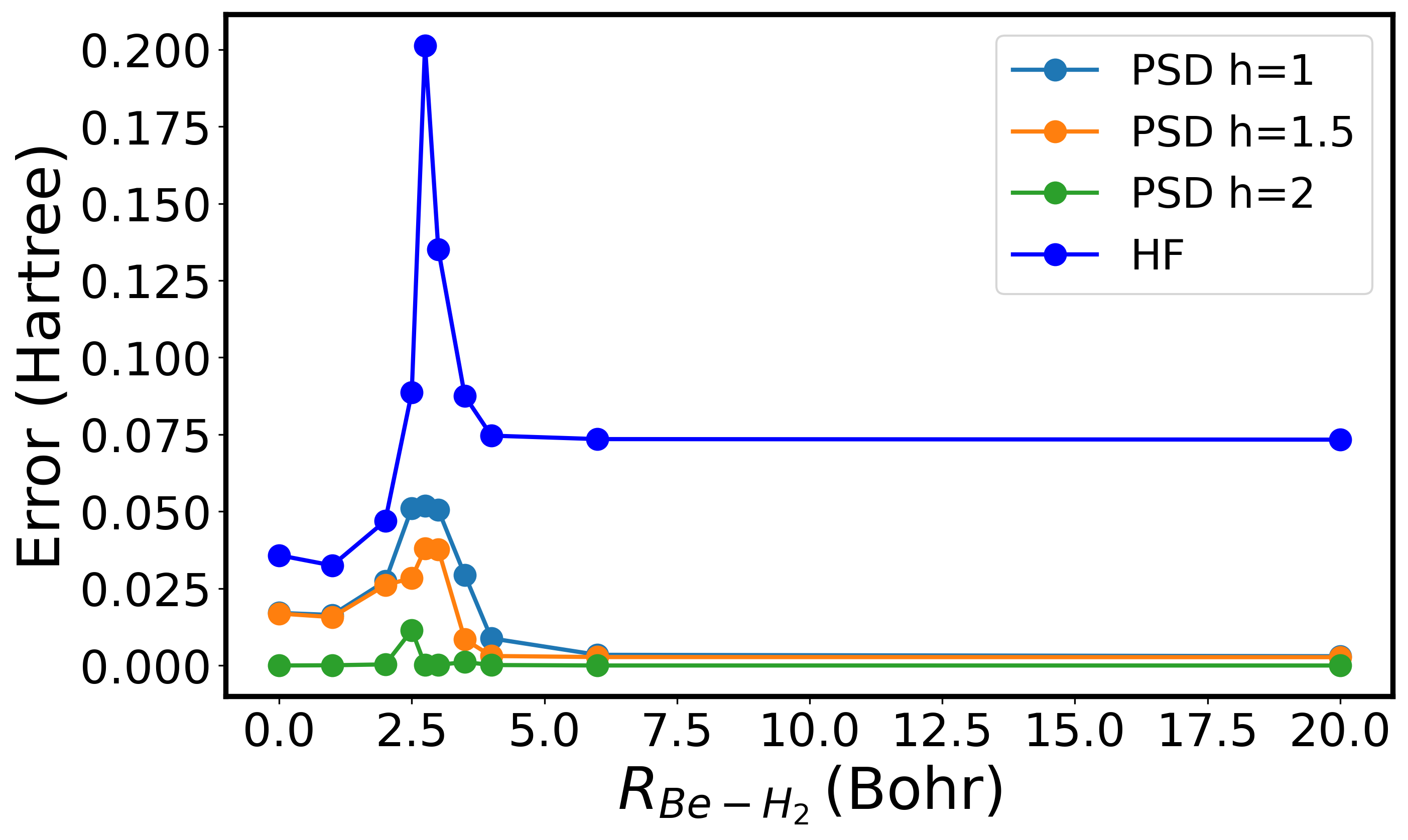} 
        \label{fig:image3}
    \end{subfigure}
    \begin{subfigure}[b]{0.45\textwidth} 
        \centering
        \includegraphics[scale=0.335]{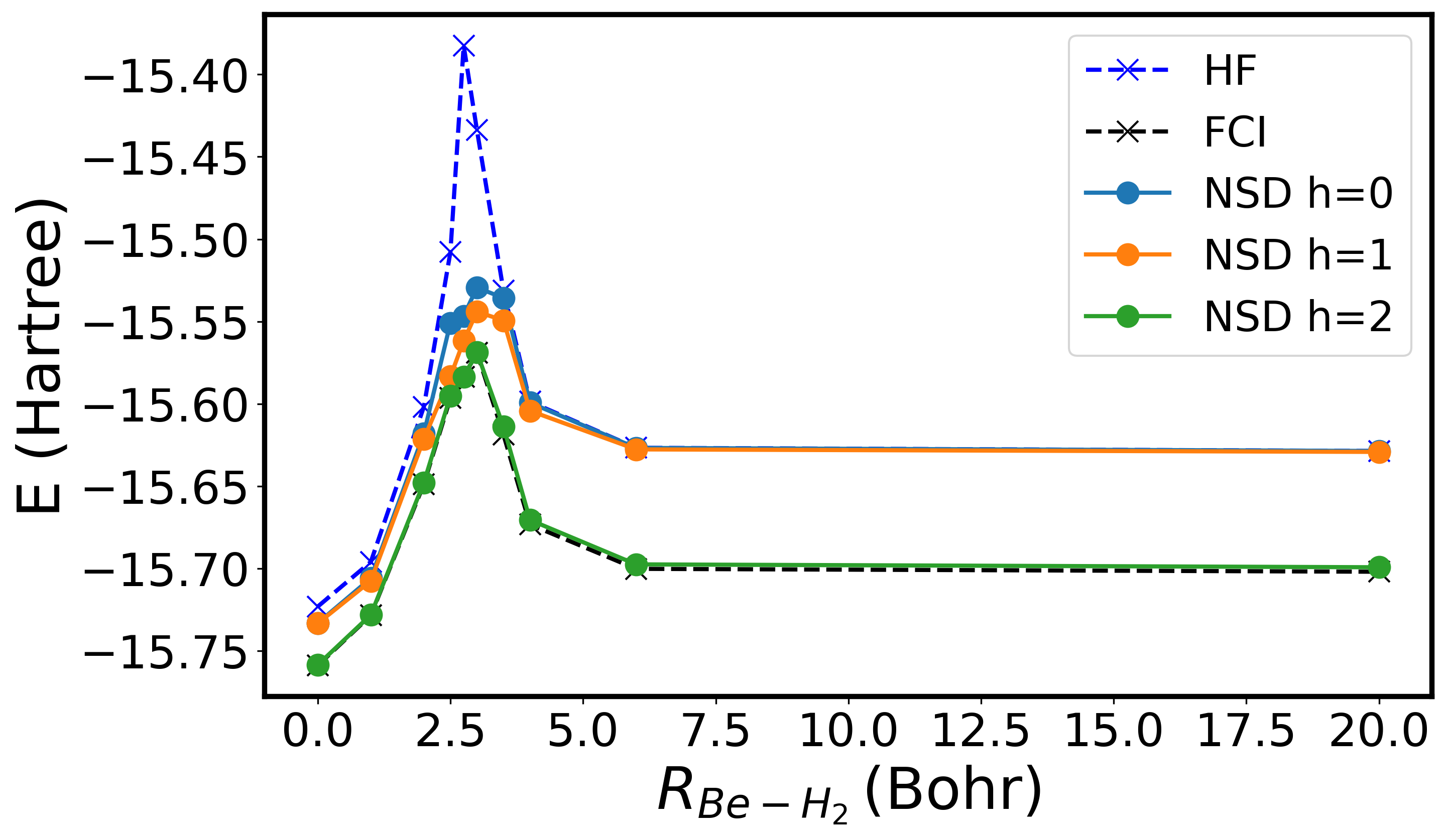} 
        \label{fig:image4}
    \end{subfigure}
    \hfill
    \begin{subfigure}[b]{0.45\textwidth} 
        \centering
        \includegraphics[scale=0.335]{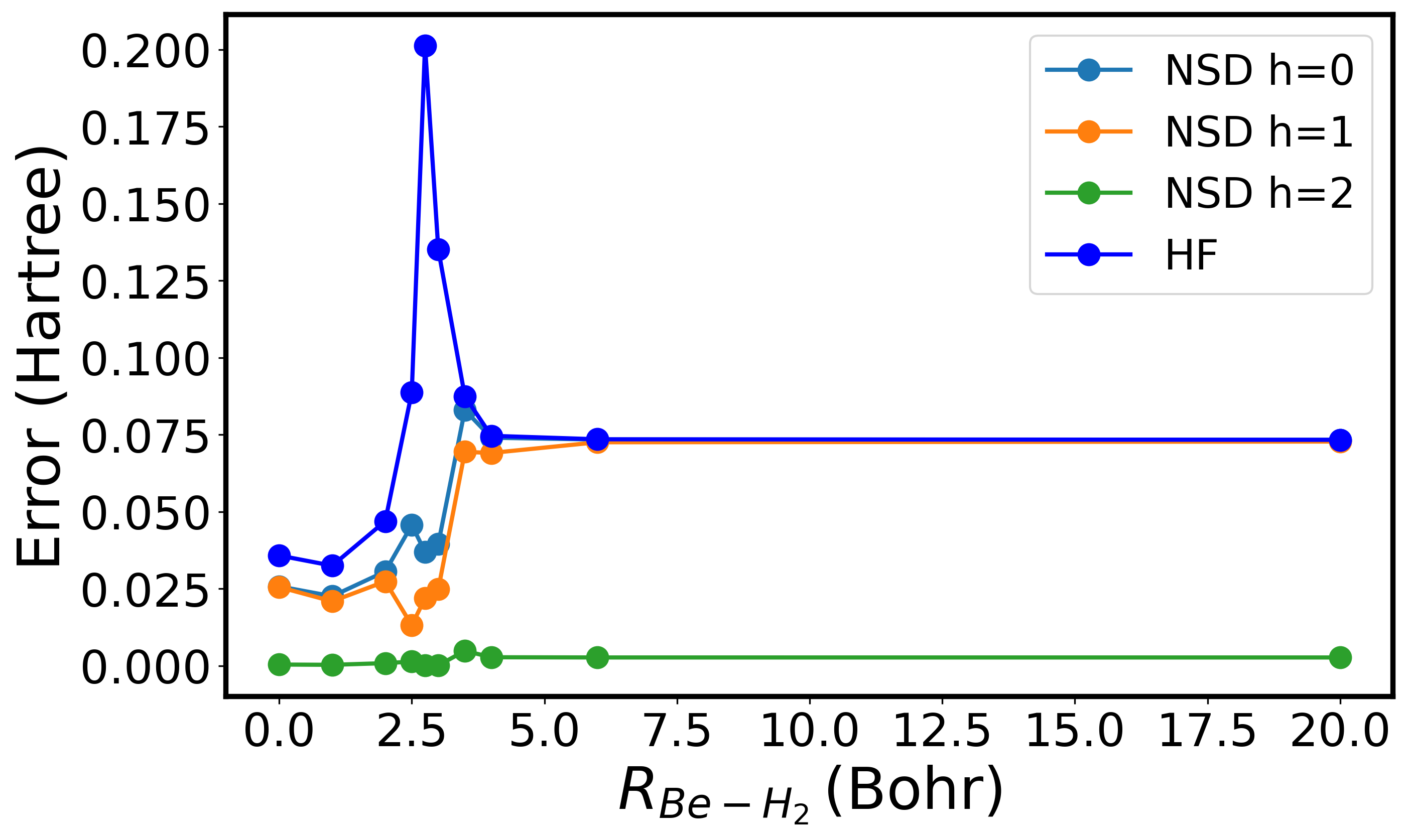} 
        \label{fig:image5}
    \end{subfigure}
    \begin{subfigure}[b]{0.45\textwidth} 
        \centering
        \includegraphics[scale=0.335]{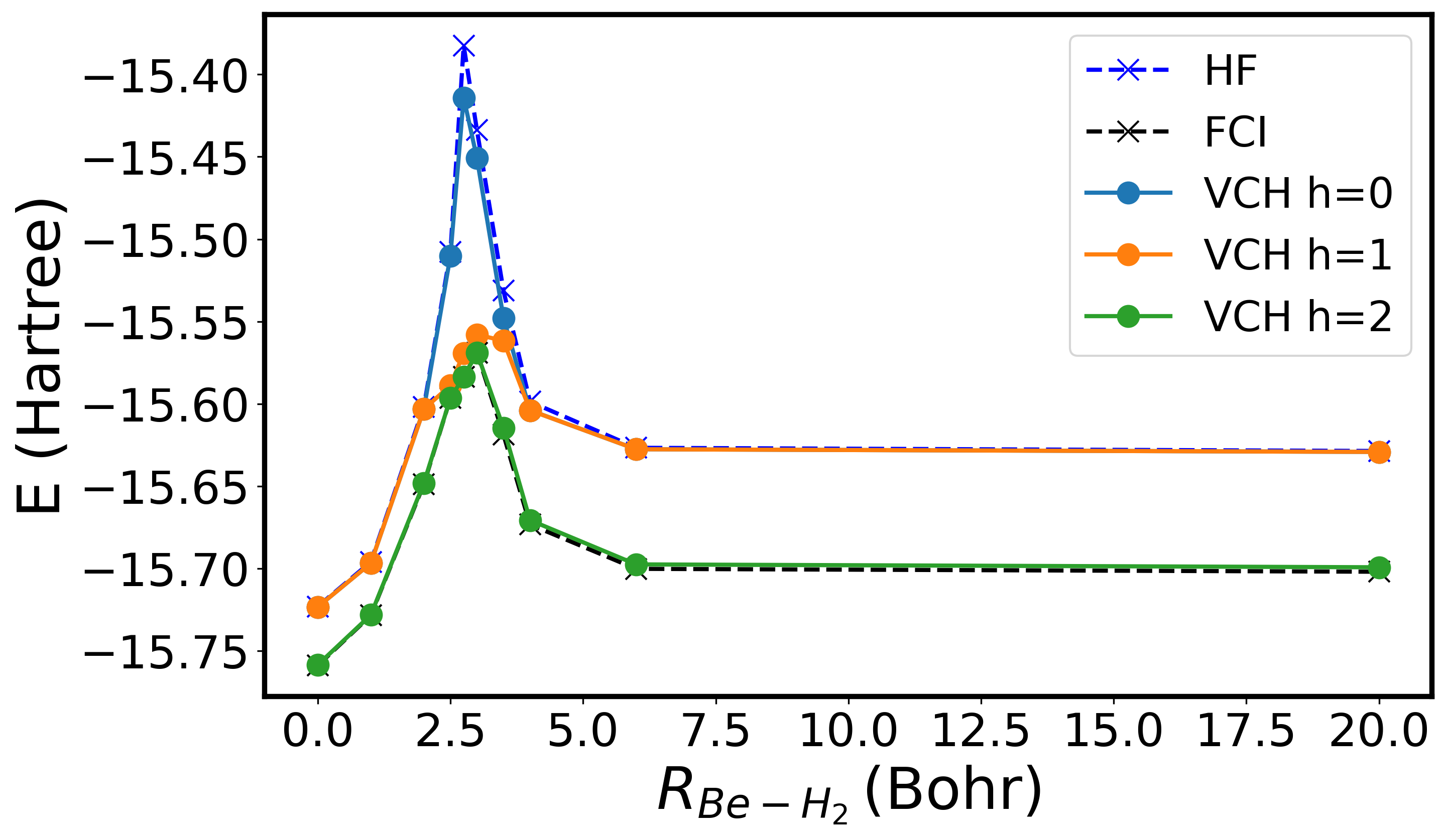} 
        \label{fig:image6}
    \end{subfigure}
    \hfill
    \begin{subfigure}[b]{0.45\textwidth} 
        \centering
        \includegraphics[scale=0.335]{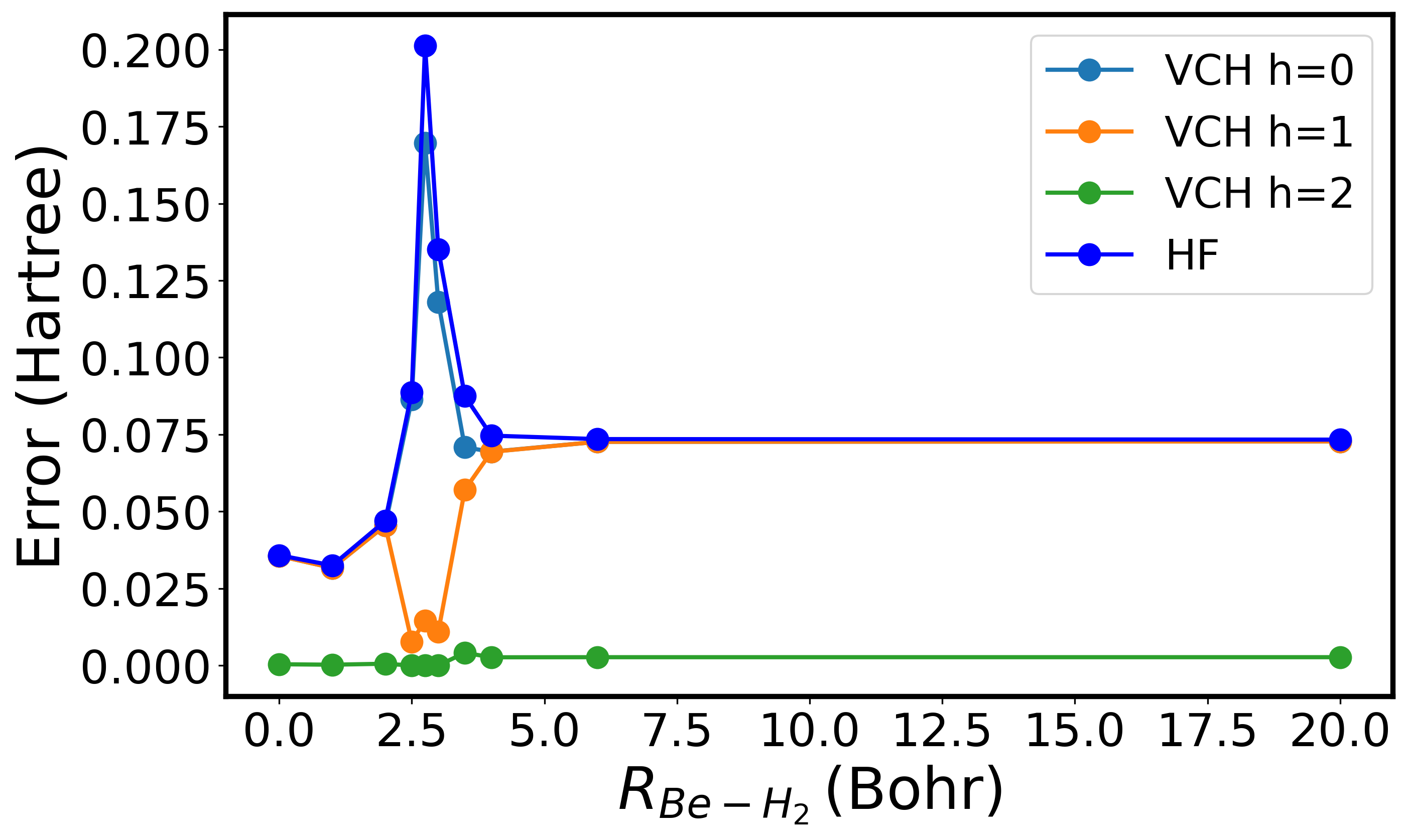} 
        \label{fig:image7}
    \end{subfigure}
    \begin{subfigure}[b]{0.45\textwidth} 
        \centering
        \includegraphics[scale=0.335]{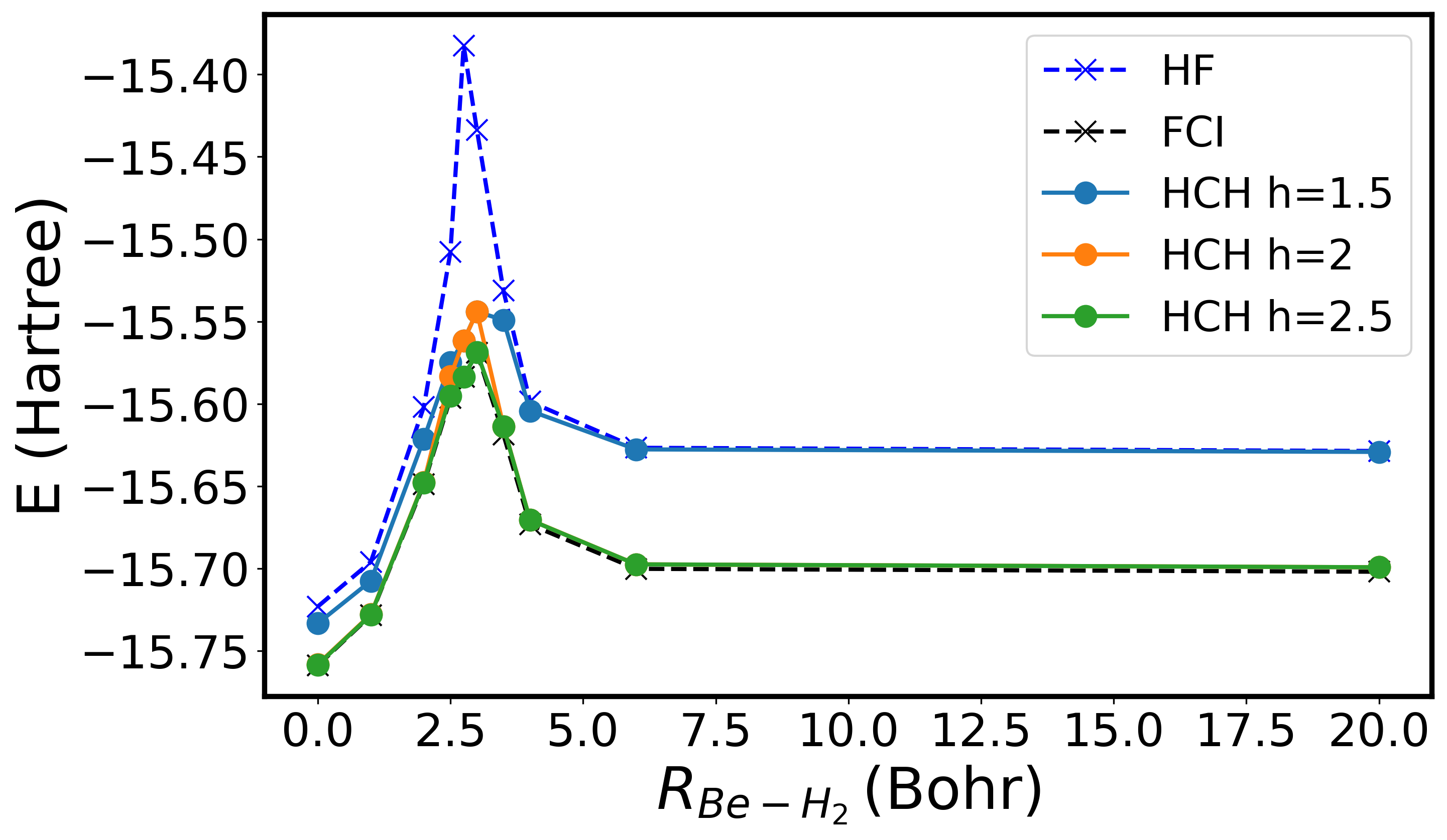} 
        \label{fig:image8}
    \end{subfigure}
    \hfill
    \begin{subfigure}[b]{0.45\textwidth} 
        \centering
        \includegraphics[scale=0.335]{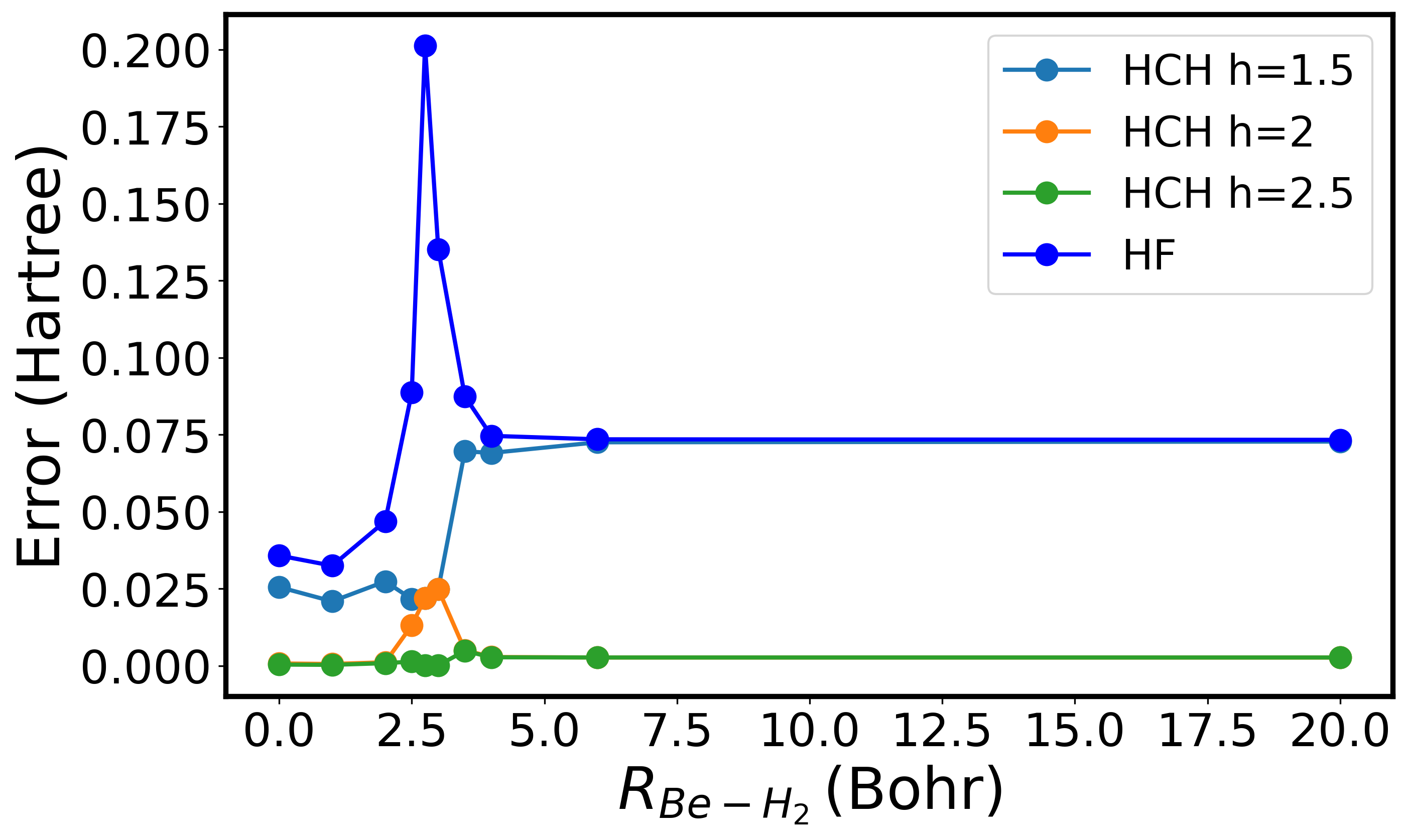} 
        \label{fig:image9}
    \end{subfigure}
\caption{    Potential energy curves for C$_{2v}$ insertion of Be into \ce{H2}, 
    showing total energies and errors relative to FCI, computed with 
    ehCI schemes (PSD, NSD, VCH, HCH) in the STO-6G basis. Curves for different hierarchy 
    levels (h-values) are indicated in the legends. }
\label{fig:results-BeH2}
\end{figure}

\autoref{fig:results-BeH2} presents total electronic energies (in Hartree) and corresponding errors with respect to FCI \[\Delta E(R)\;=\;E_{\mathrm{ehCI}}(R)\;-\;E_{\mathrm{FCI}}(R)\], versus the perpendicular \ce{Be}–\ce{H2} distance $R_{\ce{BeH2}}$ for all four ehCI schemes.
PSD partitioning ($\alpha_{1}=0.5$, $\alpha_{2}=0.25$) with increasing hierarchy effectively reproduces the FCI curve at both short and long distances, with only a slight upward shift in the mid‐range (2.5–3.0 Bohr) at lower hierarchy.
Because the magnitude of $\alpha_{2}$ (0.25) is relatively small, PSD partition includes seniority‐2 and seniority‐4 determinants at a relatively higher hierarchy number, which increases the total number of configurations and captures the multireference character under strong static correlation.
NSD partition ($\alpha_{1}=1$, $\alpha_{2}=-0.5$) deviates largely with respect to FCI at long distances but rejoins the FCI curve at a higher hierarchy.
In this case, the negative seniority weight ($\alpha_{2}=-0.5$) postpones the inclusion of seniority‐0 determinants until a higher hierarchy number, so at moderate $h_{\text{max}}$ this partition contains fewer lower order seniority determinants than PSD, resulting in a modest mid‐range offset.
Quite similarly, VCH ($\alpha_{1}=1$, $\alpha_{2}=-1$) and HCH partition ($\alpha_{1}=1$, $\alpha_{2}=-0.25$) both exhibit more pronounced deviations from FCI around $R\approx2.5$–2.75 Bohr and essentially overlapping with HF energies at lower hierarchy.
In VCH partition, the larger but negative value of $\alpha_{2}$ (especially in VCH) further delays inclusion of important seniority‐0 and excitation-1 determinants, so at each hierarchy number these partitions include a smaller determinant set of lower order \textit{e} and \textit{s} and thus recover less static correlation. 
Though the magnitude of $\alpha_2$ (seniority weight) is substantial in HCH, it encounters the seniority sectors which are inaccessible (black squares).
Both curves remain above the FCI energy until approximately $R\approx6.0$ Bohr, after which they gradually converge toward dissociation.
Nevertheless, NSD, VCH and HCH perform much better at higher hierarchies compared to PSD partitioning. 

\begin{figure}[H] 
    \centering
    \begin{subfigure}[b]{0.45\textwidth} 
        \centering
        \includegraphics[scale=0.335]{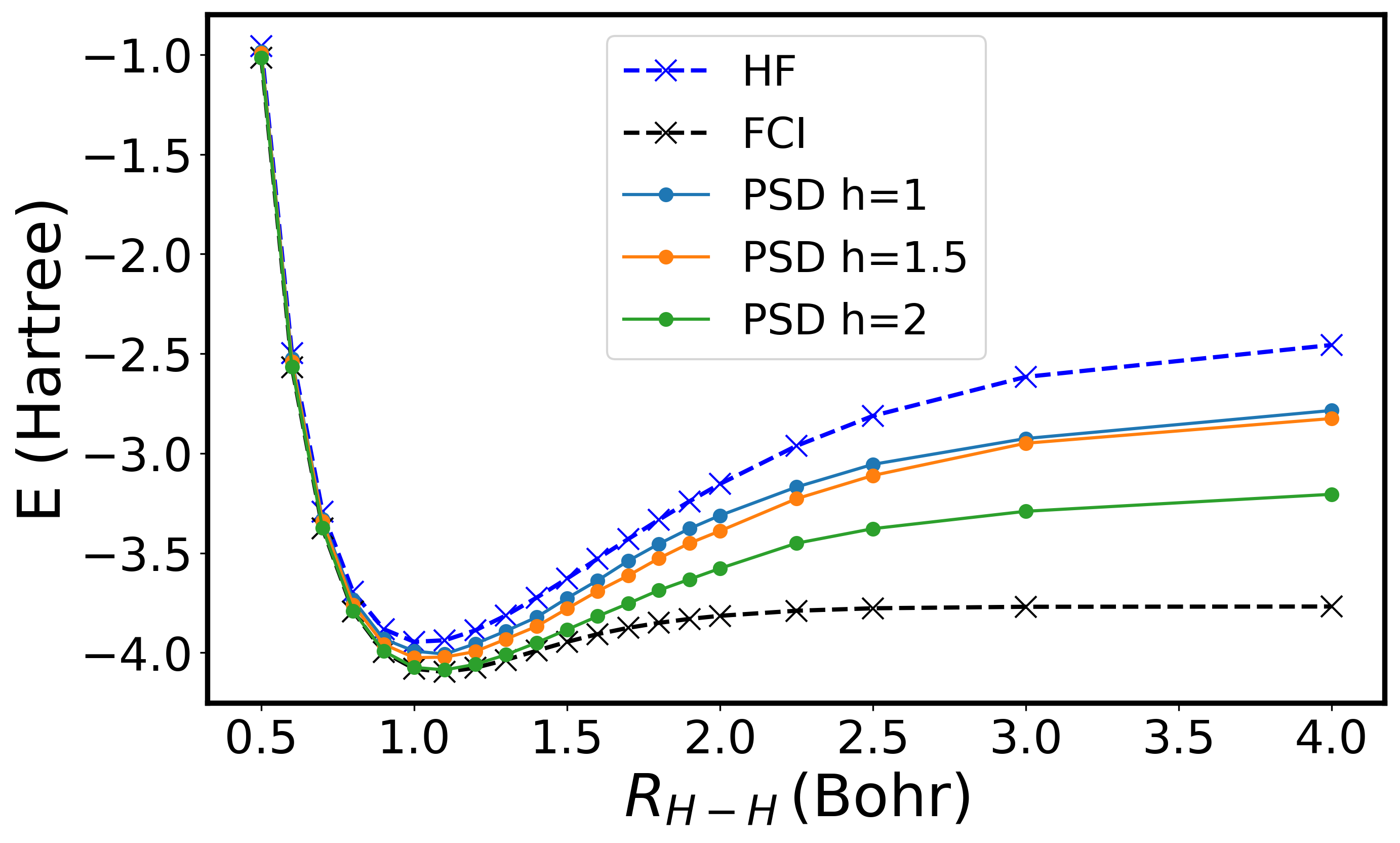} 
        \label{fig:image10}
    \end{subfigure}
    \hfill
    \begin{subfigure}[b]{0.45\textwidth} 
        \centering
        \includegraphics[scale=0.335]{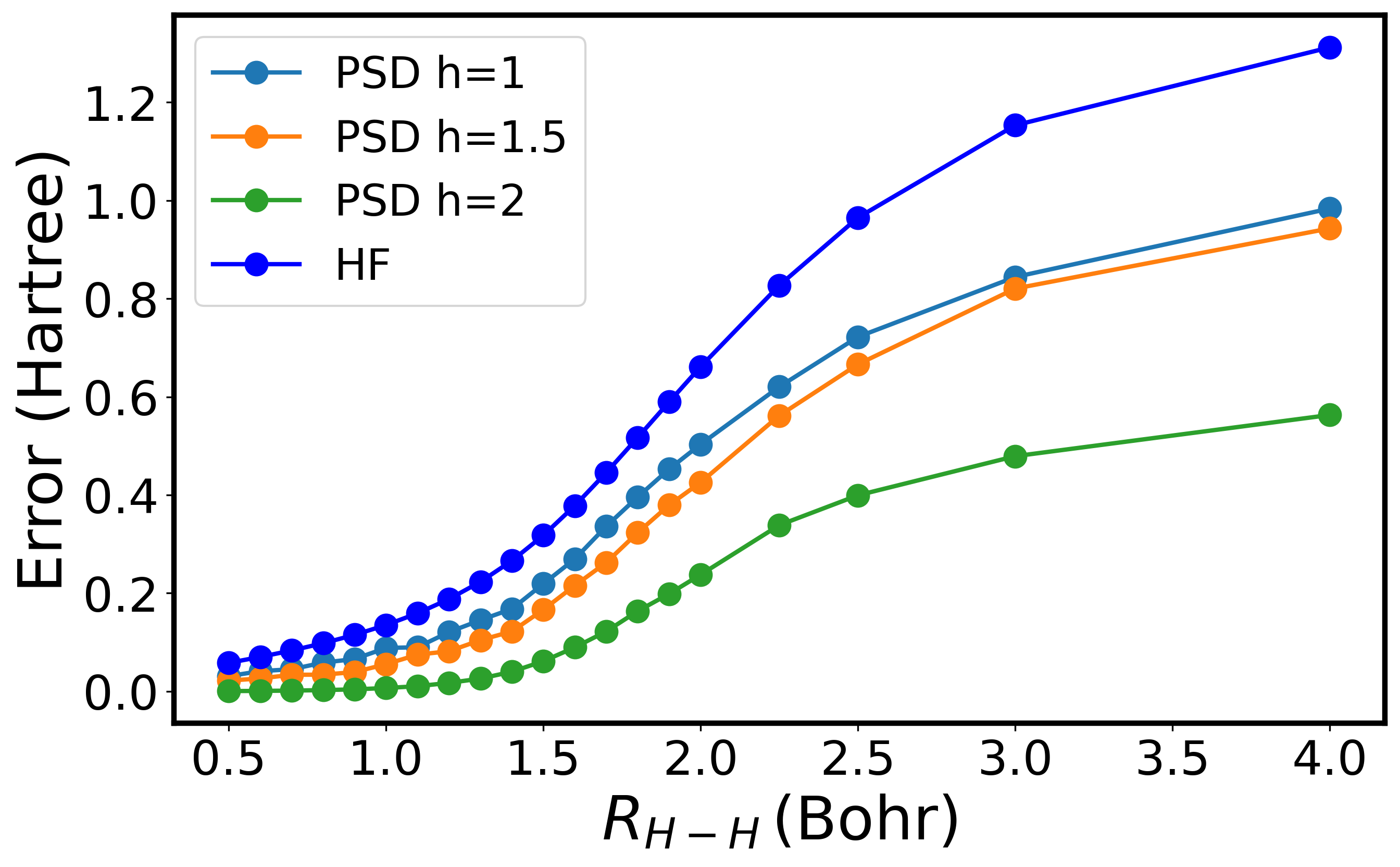} 
        \label{fig:image11}
    \end{subfigure}
    \begin{subfigure}[b]{0.45\textwidth} 
        \centering
        \includegraphics[scale=0.335]{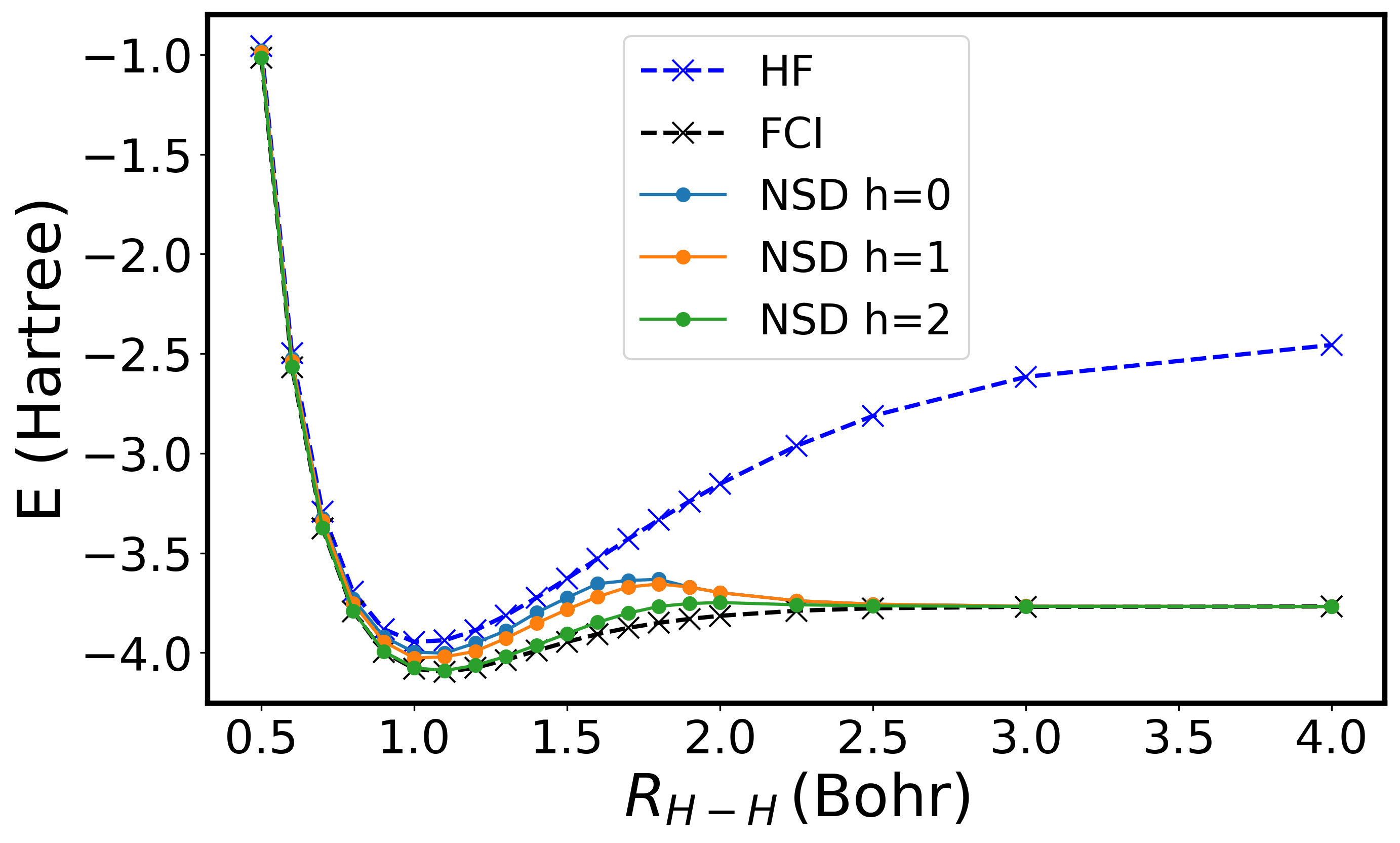} 
        \label{fig:image12}
    \end{subfigure}
    \hfill
    \begin{subfigure}[b]{0.45\textwidth} 
        \centering
        \includegraphics[scale=0.335]{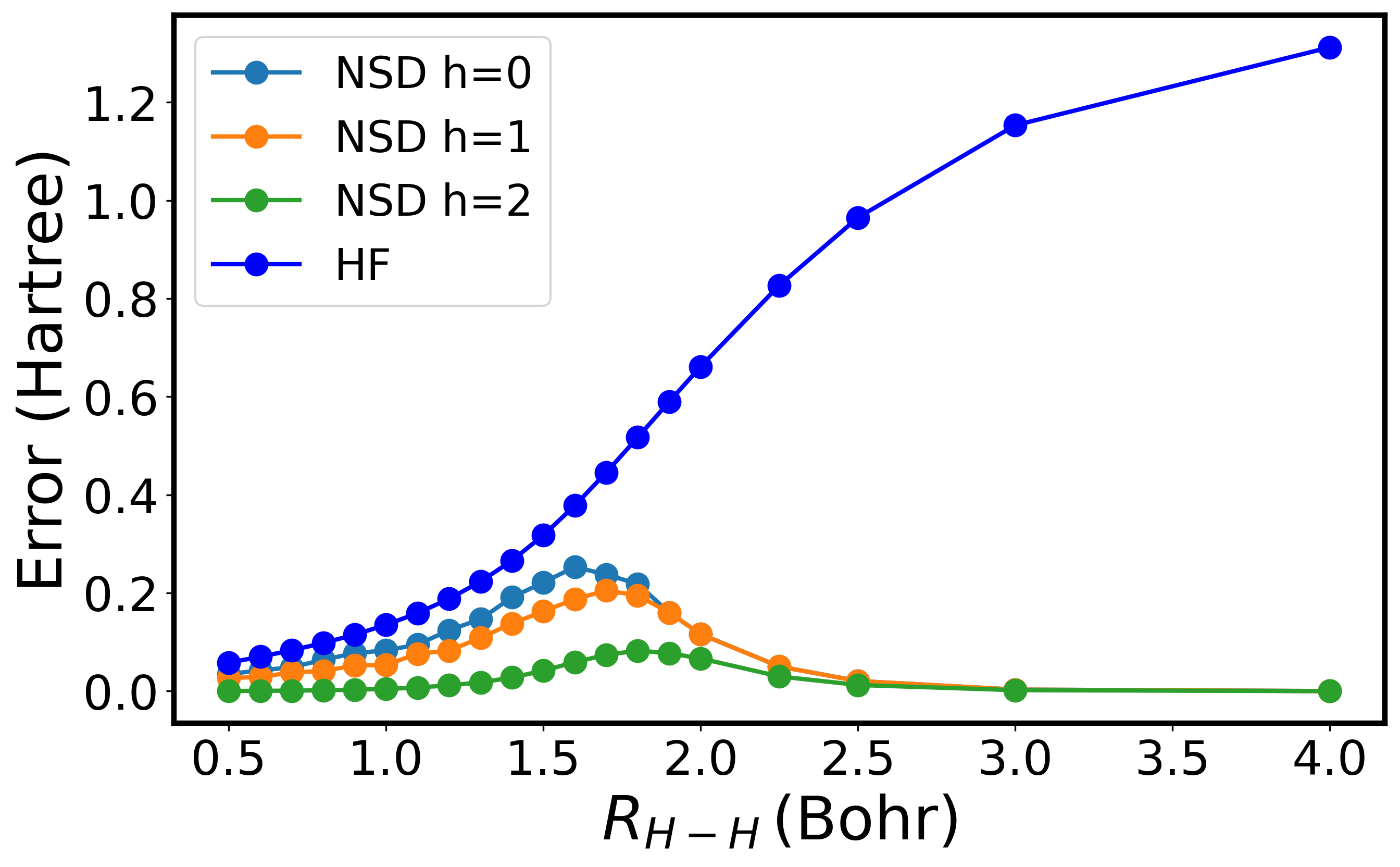} 
        \label{fig:image13}
    \end{subfigure}
    \begin{subfigure}[b]{0.45\textwidth} 
        \centering
        \includegraphics[scale=0.335]{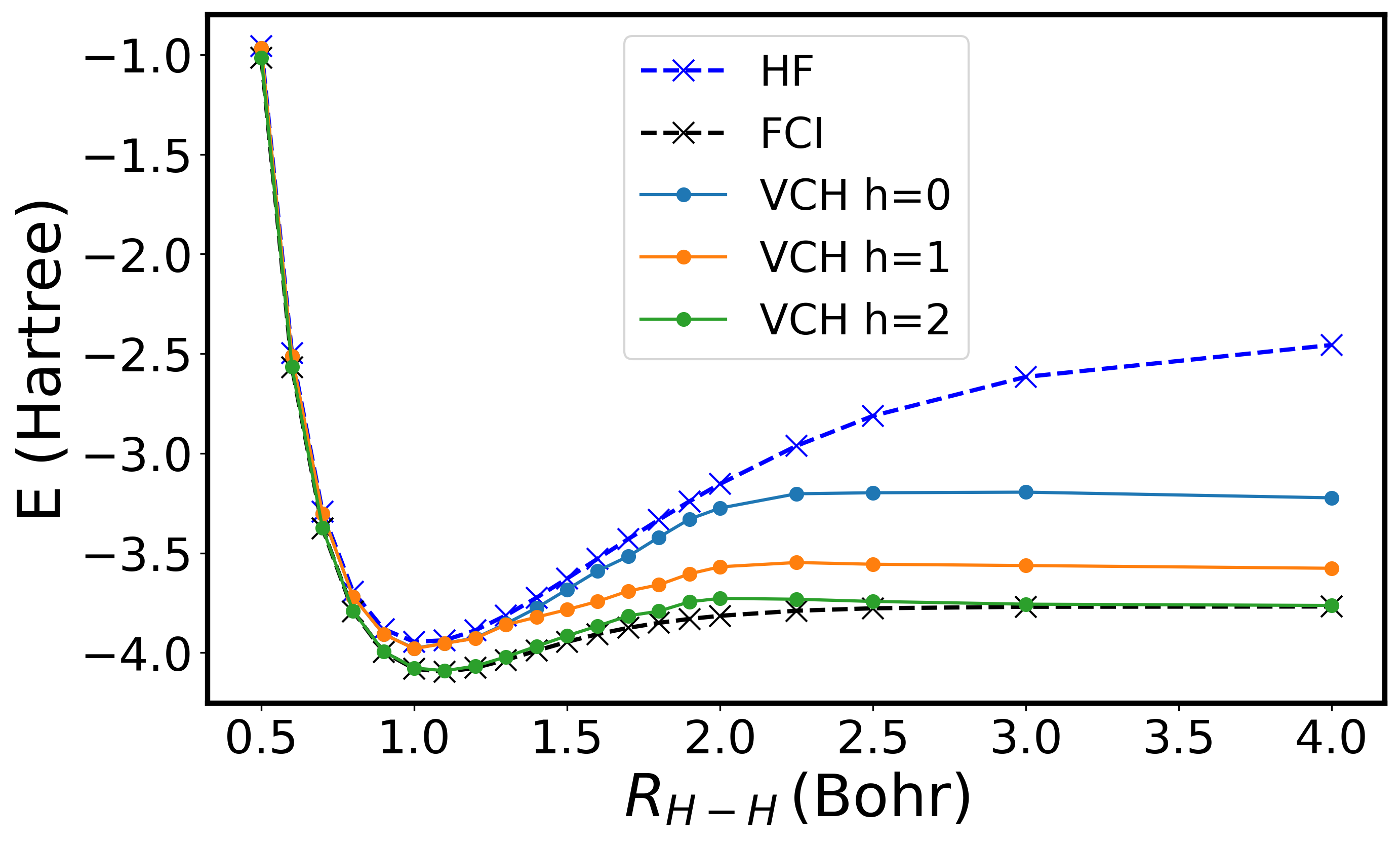} 
        \label{fig:image14}
    \end{subfigure}
    \hfill
    \begin{subfigure}[b]{0.45\textwidth} 
        \centering
        \includegraphics[scale=0.335]{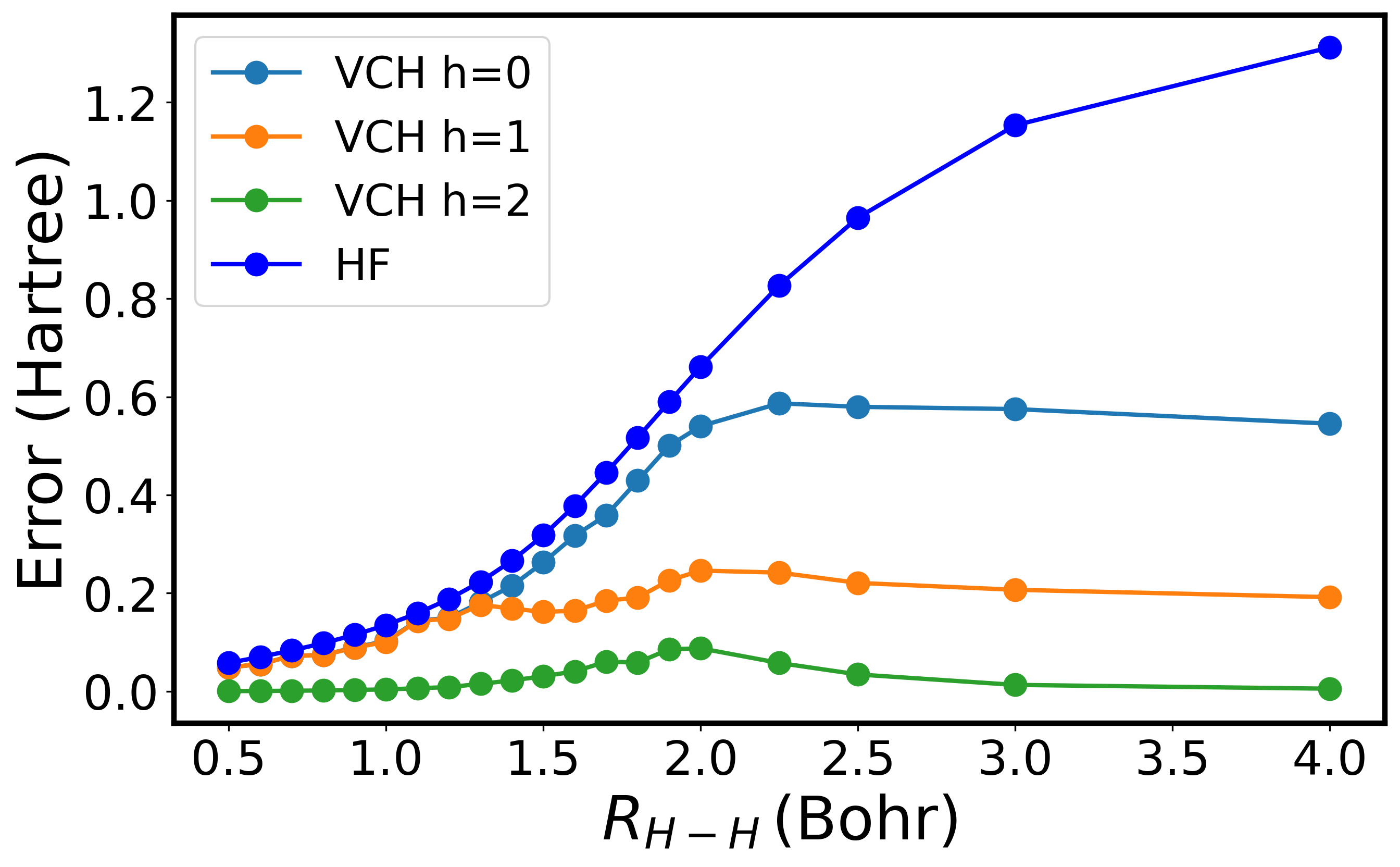} 
        \label{fig:image15}
    \end{subfigure}
    \begin{subfigure}[b]{0.45\textwidth} 
        \centering
        \includegraphics[scale=0.335]{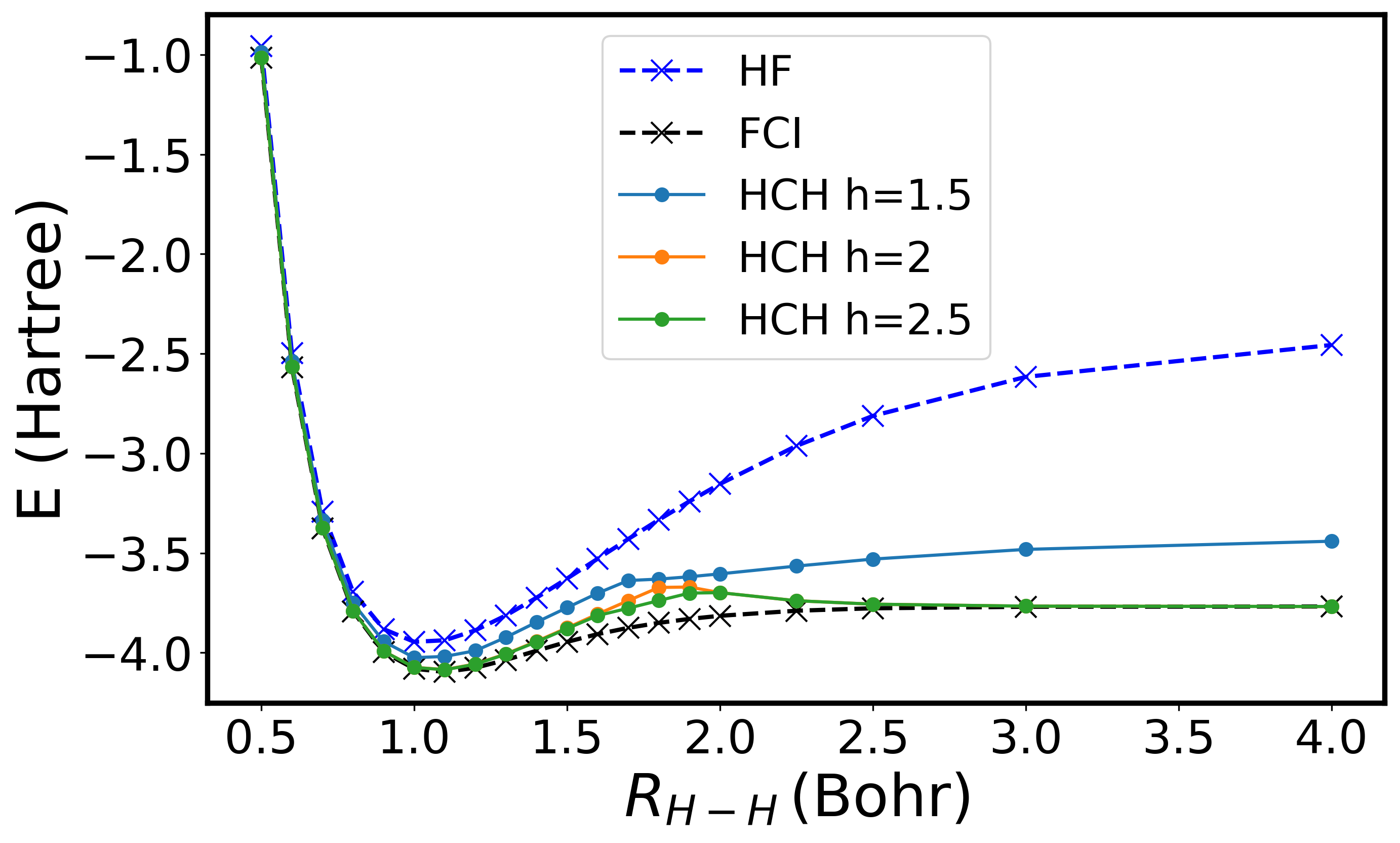} 
        \label{fig:image16}
    \end{subfigure}
    \hfill
    \begin{subfigure}[b]{0.45\textwidth} 
        \centering
        \includegraphics[scale=0.335]{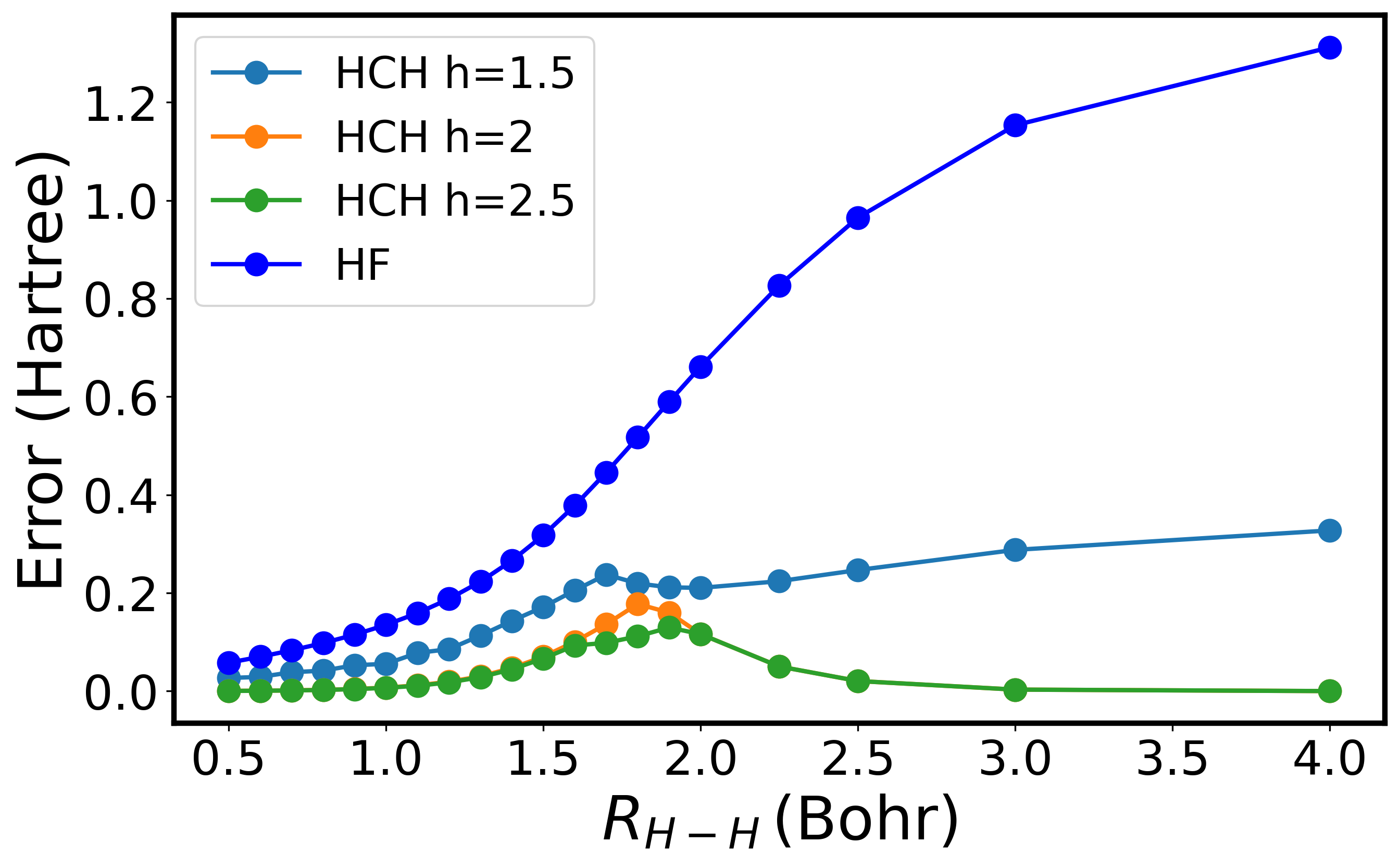} 
        \label{fig:image17}
    \end{subfigure}
    \caption{Potential energy curves of cubic \ce{H8} dissociation representing total energies and energy errors relative to FCI, computed using the ehCI methods using the STO-6G basis. Curves for different hierarchy 
    levels (h-values) are indicated in the legends. }
    \label{fig:results-h8_cube}
\end{figure}
\subsection{\ce{H8} curves}
\autoref{fig:results-h8_cube} presents the total electronic energy of the cubic \ce{H8} cluster (in Hartree) as a function of cube edge length $L$ (0.5–4.0 Bohr), comparing the four ehCI partitions (PSD, NSD, VCH, HCH) to FCI.
It is also important to note that as the electron count increases from six in \ce{BeH2} to eight in the \ce{H8} cube, the number of Slater determinants for a given excitation degree $e$ and seniority $s$ grows combinatorially.
For an $N$-electron system, the number of possible excitations of degree $e$ can be determined by binomial coefficients $\binom{N}{e}$, and the number of seniority-$s$ configurations by additional combinatorial factors.
Thus, at the same hierarchy cutoff, the eight-electron \ce{H8} system possesses a vastly larger determinant space than the six-electron \ce{BeH2} system, which in turn produces a different pattern of results for equivalent levels of ($e$,$s$) pairs.
For e.g. NSD for h=0 has 593 determinants for \ce{BeH2} and 1971 determinants for \ce{H8} system. 

PSD partitioning, with $h=1$, closely follows the FCI curve over most of the range but exhibits a noticeable upward deviation in the bond dissociation regime.
As the hierarchy increases, the energy error decreases.
Partitions NSD, VCH, and HCH similarly overestimate the energy in this static‐correlation regime for $R_{H-H} > 2.0$ Bohr, but they perform better compared to PSD for a higher hierarchy since the number of determinants added for an increasing hierarchy number in these partitions is an order of magnitude greater compared to PSD. 

\begin{figure}[H] 
    \centering
    \begin{subfigure}[b]{0.45\textwidth} 
        \centering
        \includegraphics[scale=0.335]{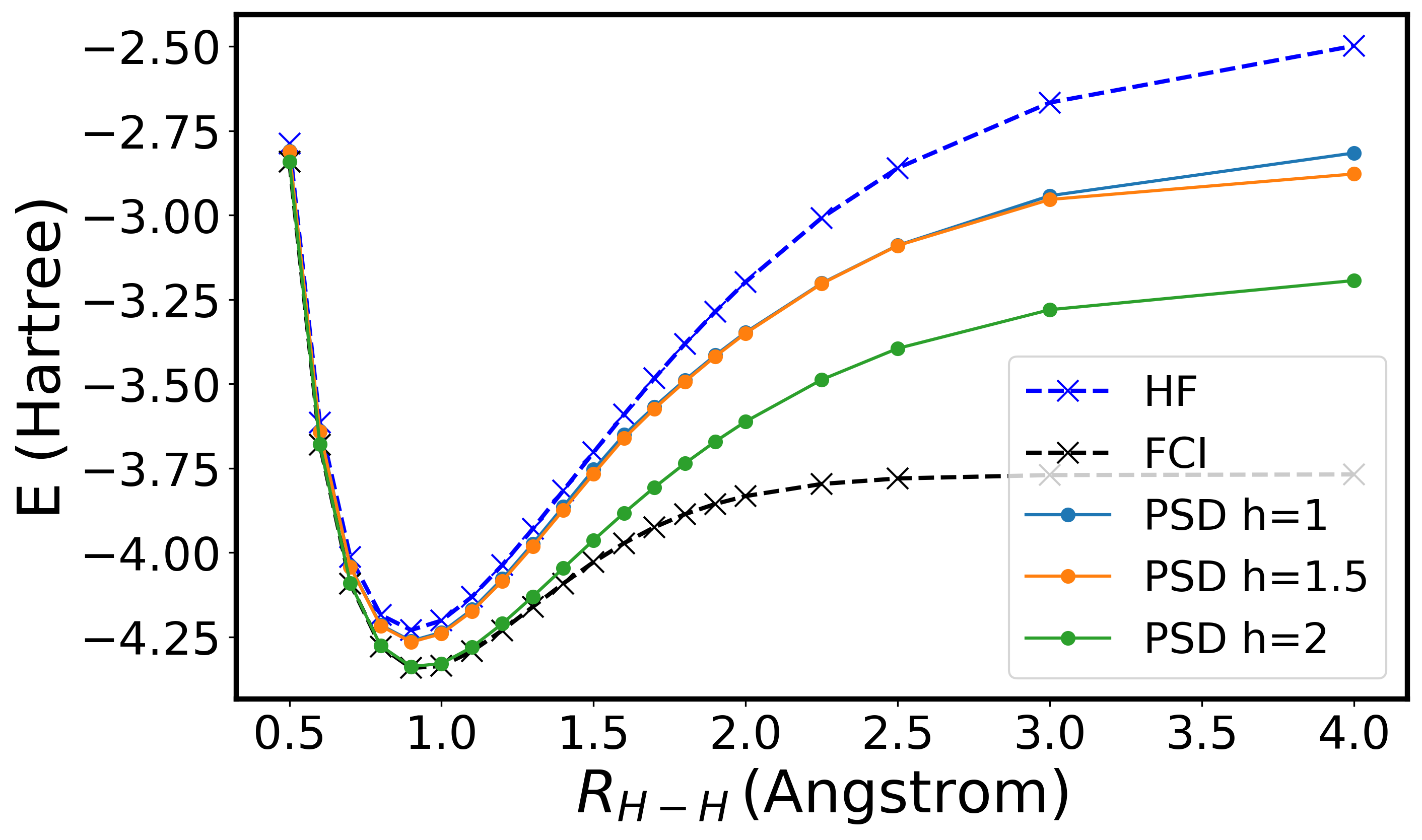} 
        \label{fig:image18}
    \end{subfigure}
    \hfill
    \begin{subfigure}[b]{0.45\textwidth} 
        \centering
        \includegraphics[scale=0.335]{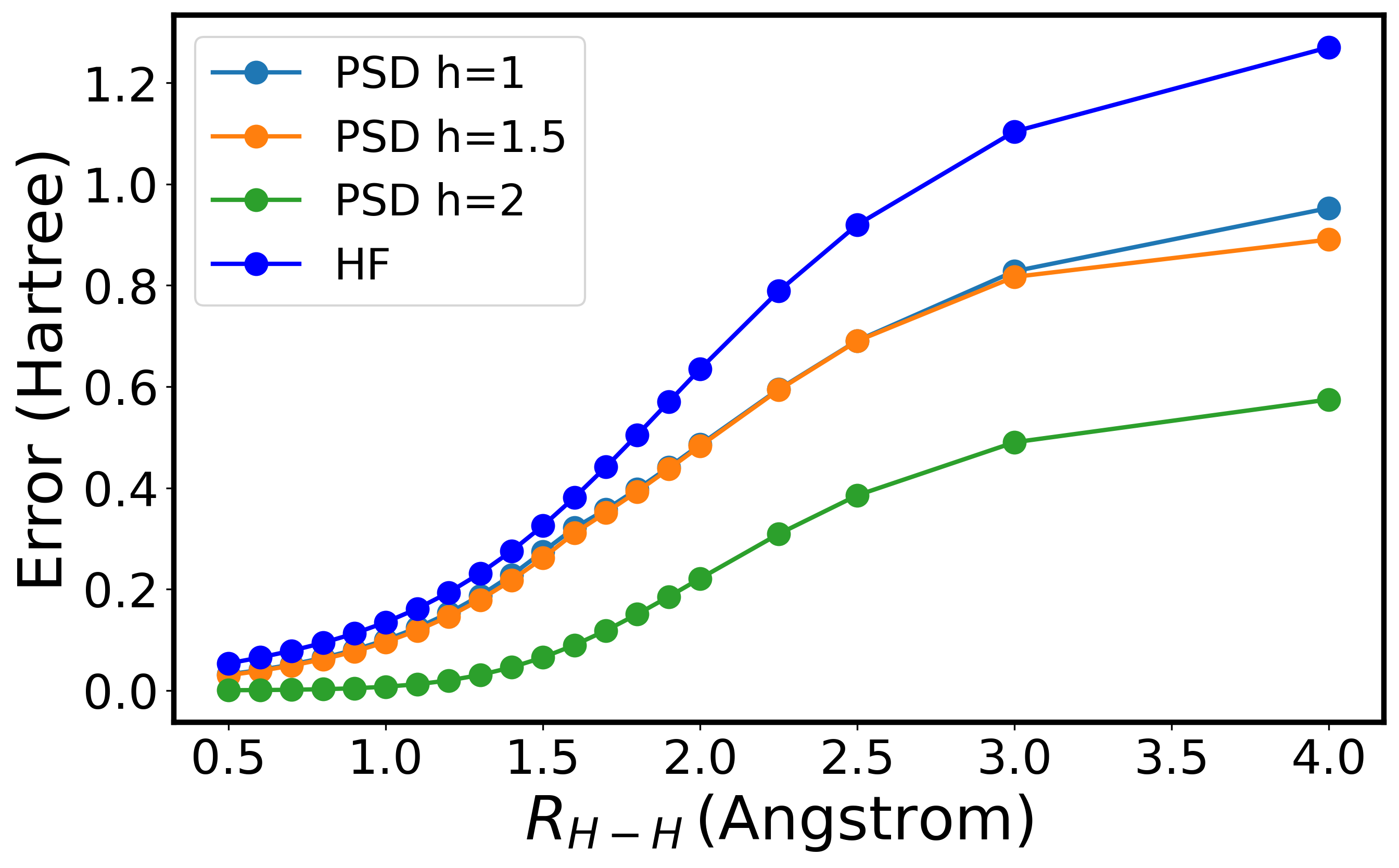} 
        \label{fig:image19}
    \end{subfigure}
    \begin{subfigure}[b]{0.45\textwidth} 
        \centering
        \includegraphics[scale=0.335]{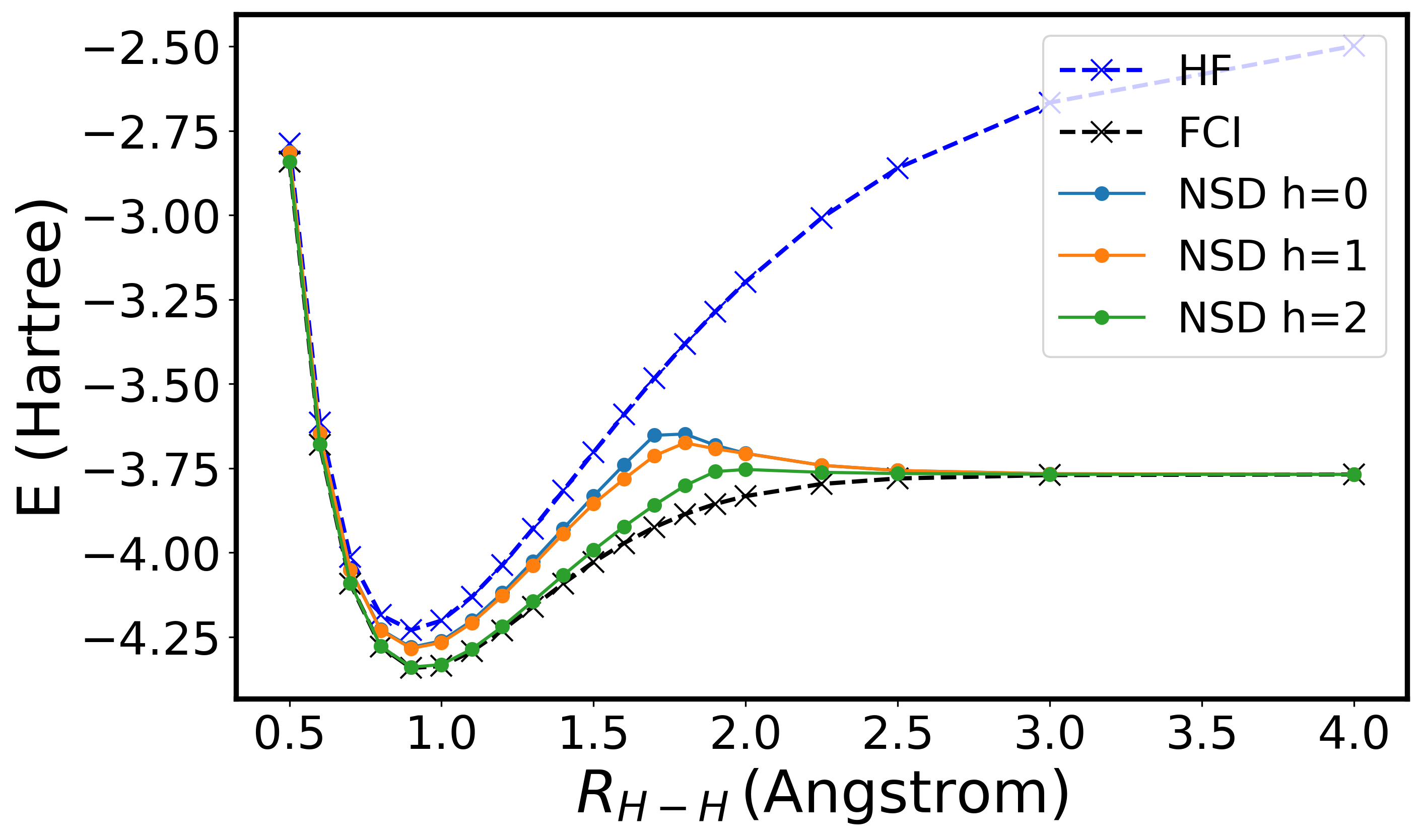} 
        \label{fig:image20}
    \end{subfigure}
    \hfill
    \begin{subfigure}[b]{0.45\textwidth} 
        \centering
        \includegraphics[scale=0.335]{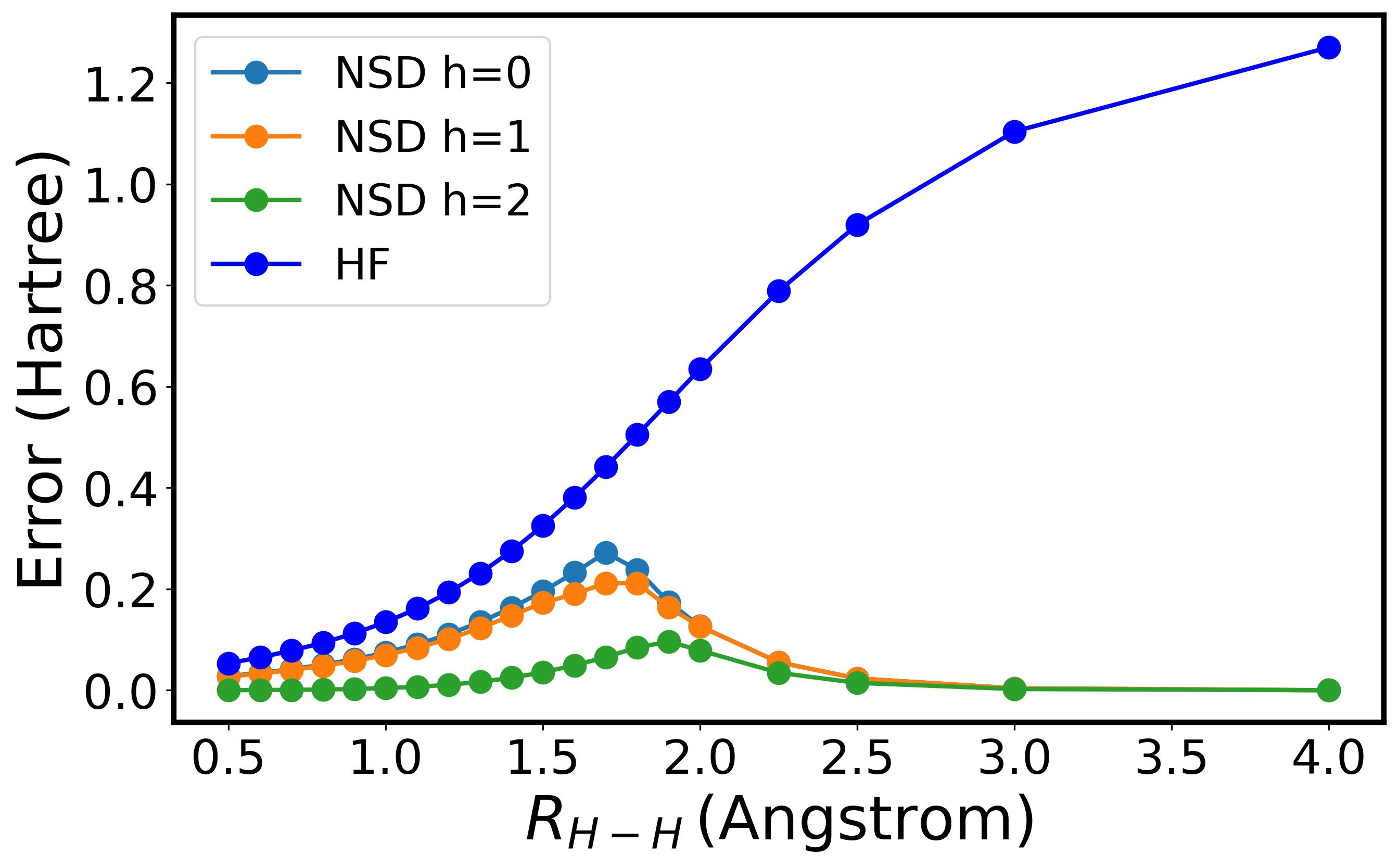} 
        \label{fig:image21}
    \end{subfigure}
    \begin{subfigure}[b]{0.45\textwidth} 
        \centering
        \includegraphics[scale=0.335]{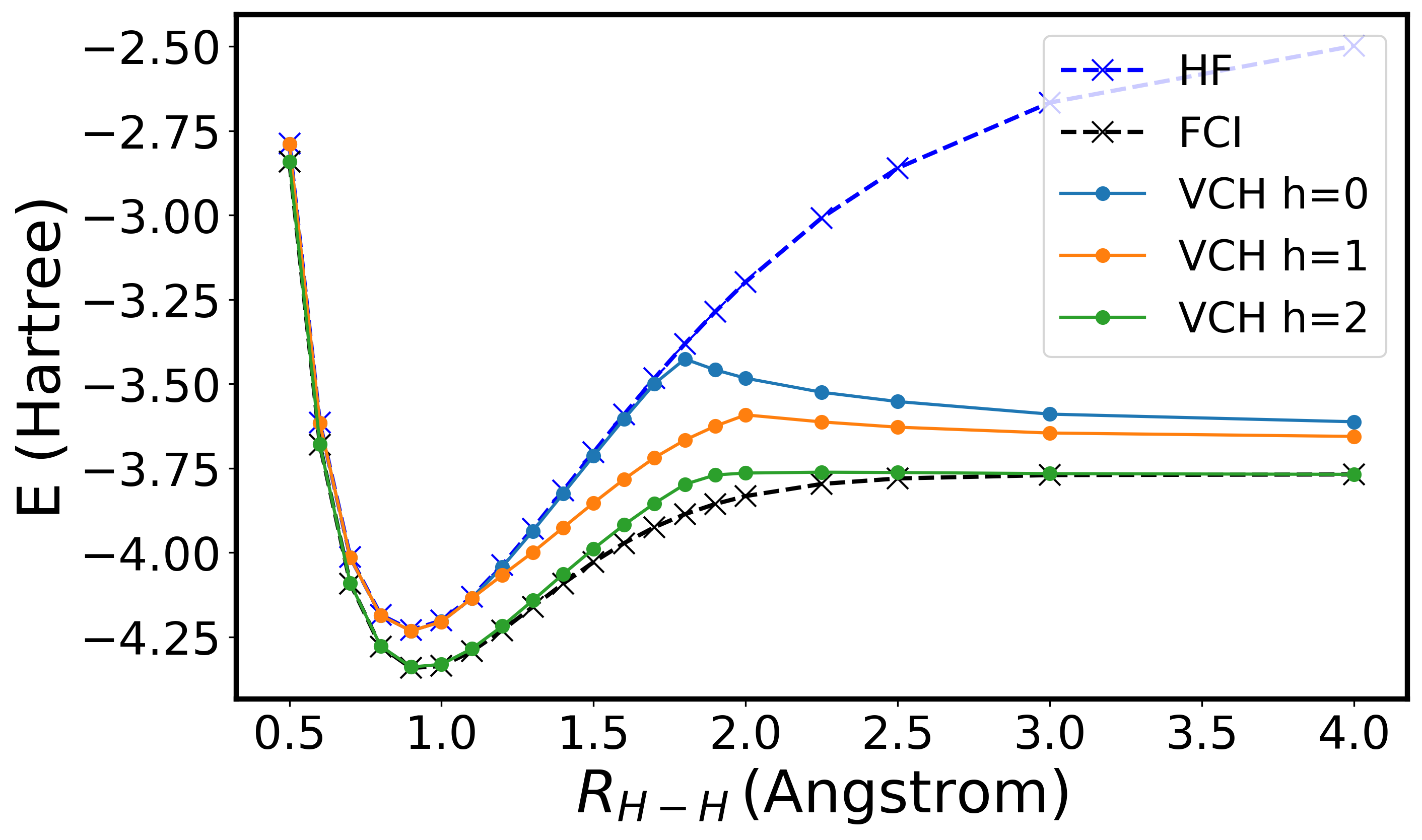} 
        \label{fig:image22}
    \end{subfigure}
    \hfill
    \begin{subfigure}[b]{0.45\textwidth} 
        \centering
        \includegraphics[scale=0.335]{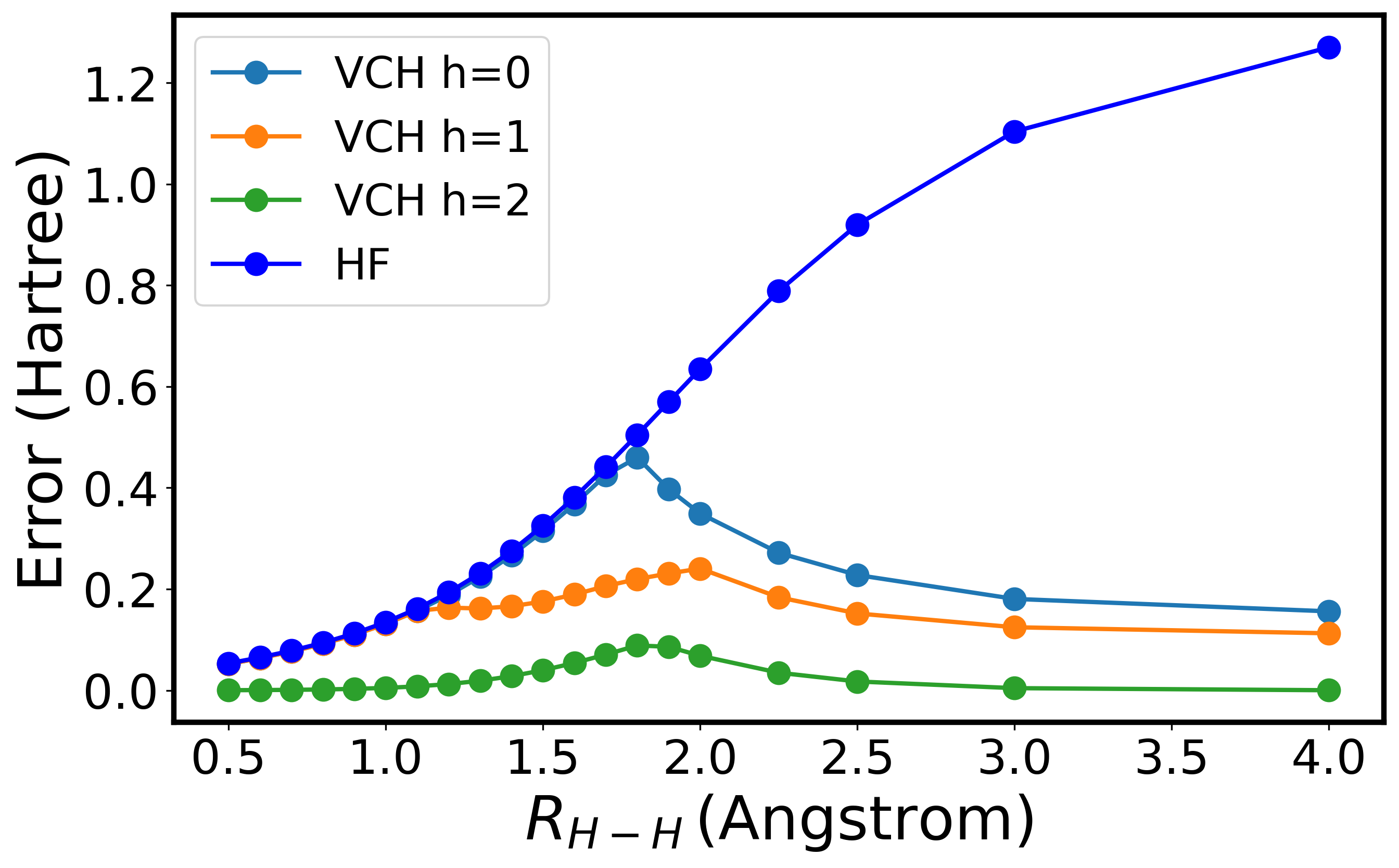} 
        \label{fig:image23}
    \end{subfigure}
    \begin{subfigure}[b]{0.45\textwidth} 
        \centering
        \includegraphics[scale=0.335]{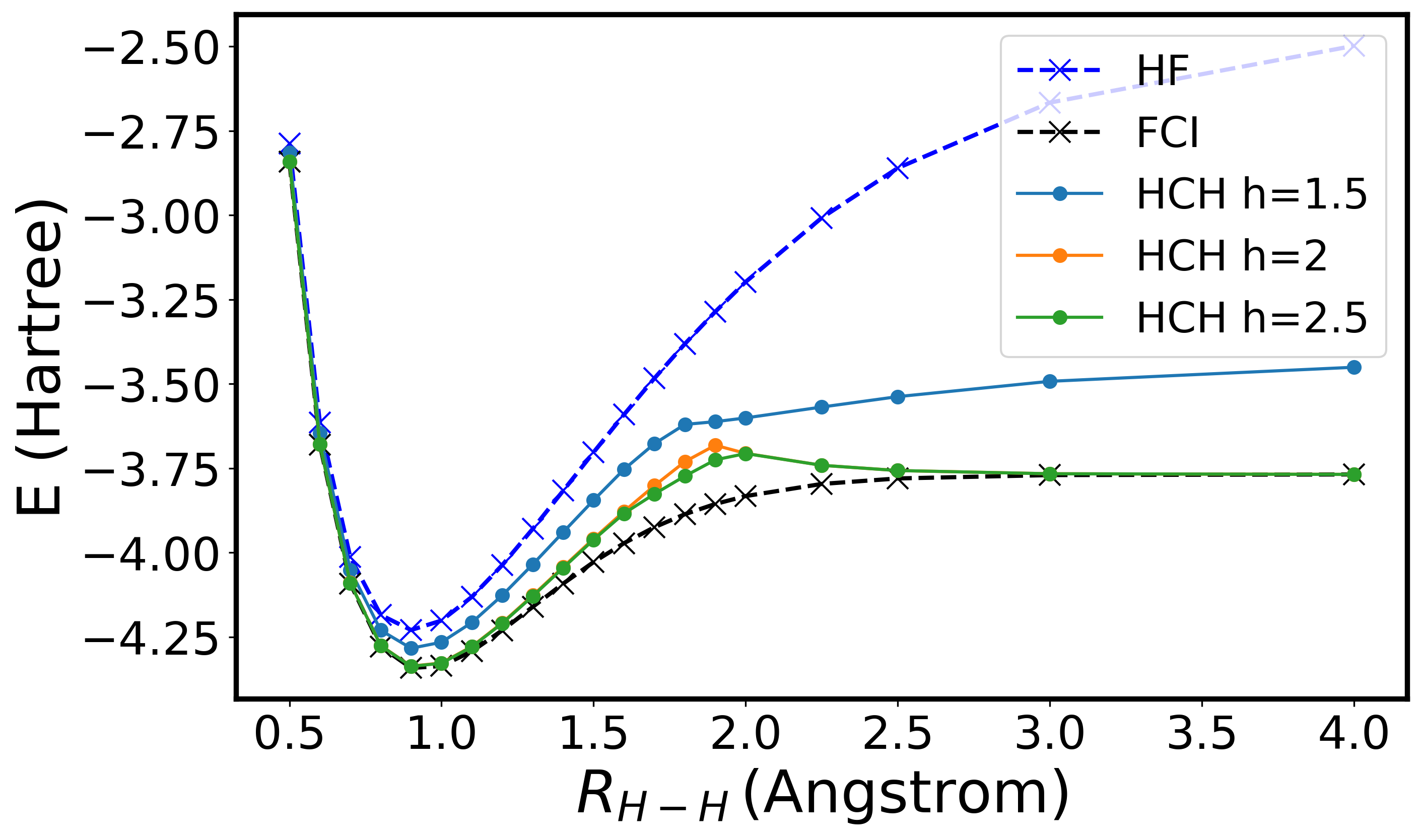} 
        \label{fig:image24}
    \end{subfigure}
    \hfill
    \begin{subfigure}[b]{0.45\textwidth} 
        \centering
        \includegraphics[scale=0.335]{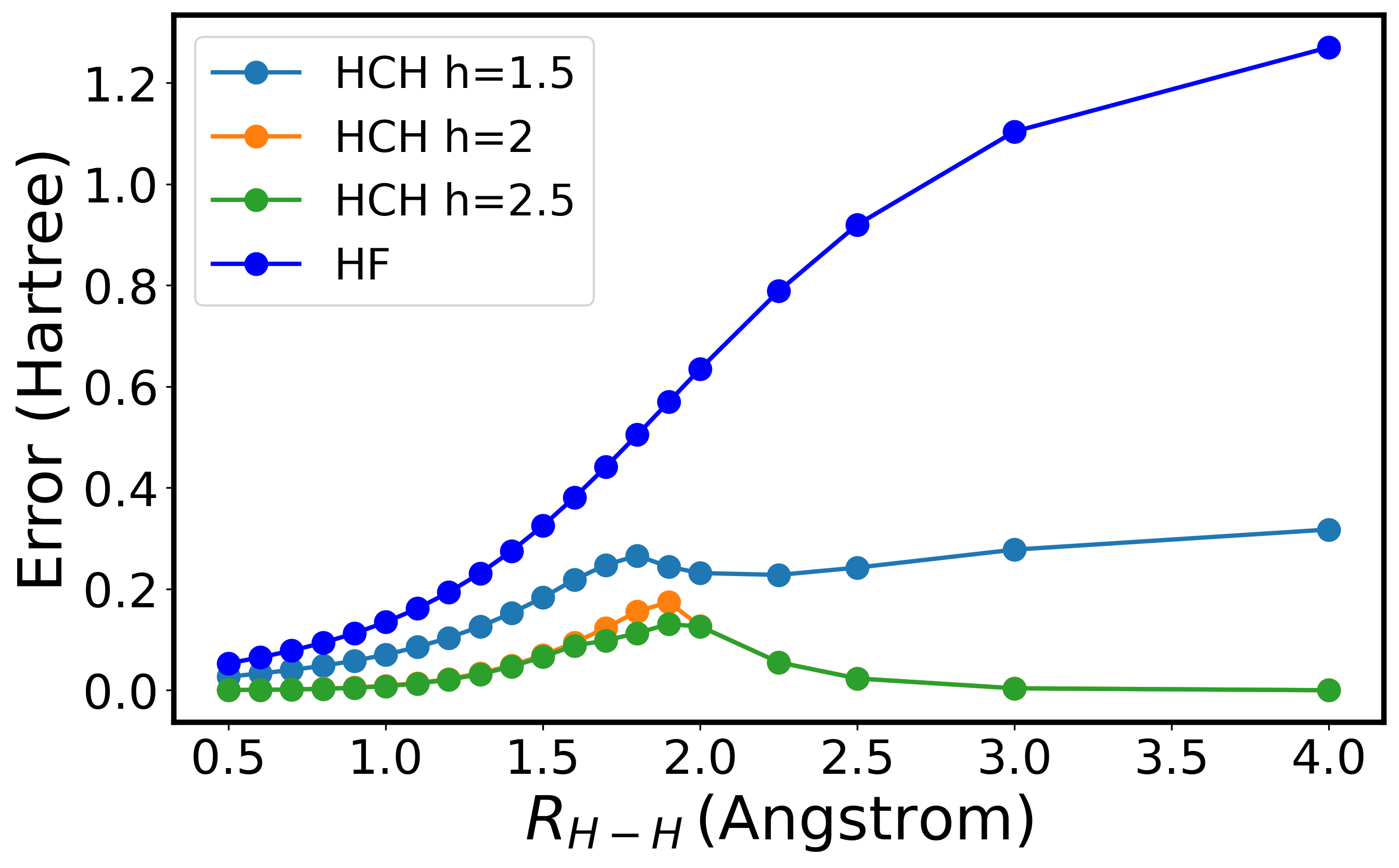} 
        \label{fig:image25}
    \end{subfigure}
    \caption{Potential energy curves of linear \ce{H8} dissociation representing total energies and and energy errors relative to FCI, computed using the ehCI methods using STO-6G basis. Curves for different hierarchy 
    levels (h-values) are indicated in the legends. 
}
\label{fig:results-h8_linear}
\end{figure}

\autoref{fig:results-h8_linear} illustrates the potential energy curves for the symmetric dissociation of linear \ce{H8}, plotted in Hartree, together with the energy deviations relative to FCI.
The overall behavior closely parallels that observed for the cubic \ce{H8} cluster.
At low hierarchy cutoffs in NSD and VCH partitioning, the potential energy profiles exhibit cusp-like features.
These sharp features can give the false impression of a crossing between two potential energy surfaces.
However, as the hierarchy cutoff increases and additional (\textit{e}, \textit{s}) pairs are added, these discontinuities are progressively eliminated.
The energy profiles then become smoother and show systematic convergence.
The ordering of maximum errors for the \ce{H8} systems (PSD $<$ HCH  $<$ NSD $<$ VCH) directly correlates with the number of determinants included at a given hierarchy number: greater $h_{\max}$ admits more seniority-2 and seniority-4 configurations, systematically improving static‐correlation recovery.
Indeed, as the hierarchy cutoff increases, all partitioning schemes converge monotonically toward FCI, underscoring the pivotal role of hierarchy number in capturing both static and dynamic correlation for the \ce{H8} cube dissociation.

\begin{figure}[H] 
    \centering
    \begin{subfigure}[b]{0.45\textwidth} 
        \centering
        \includegraphics[scale=0.335]{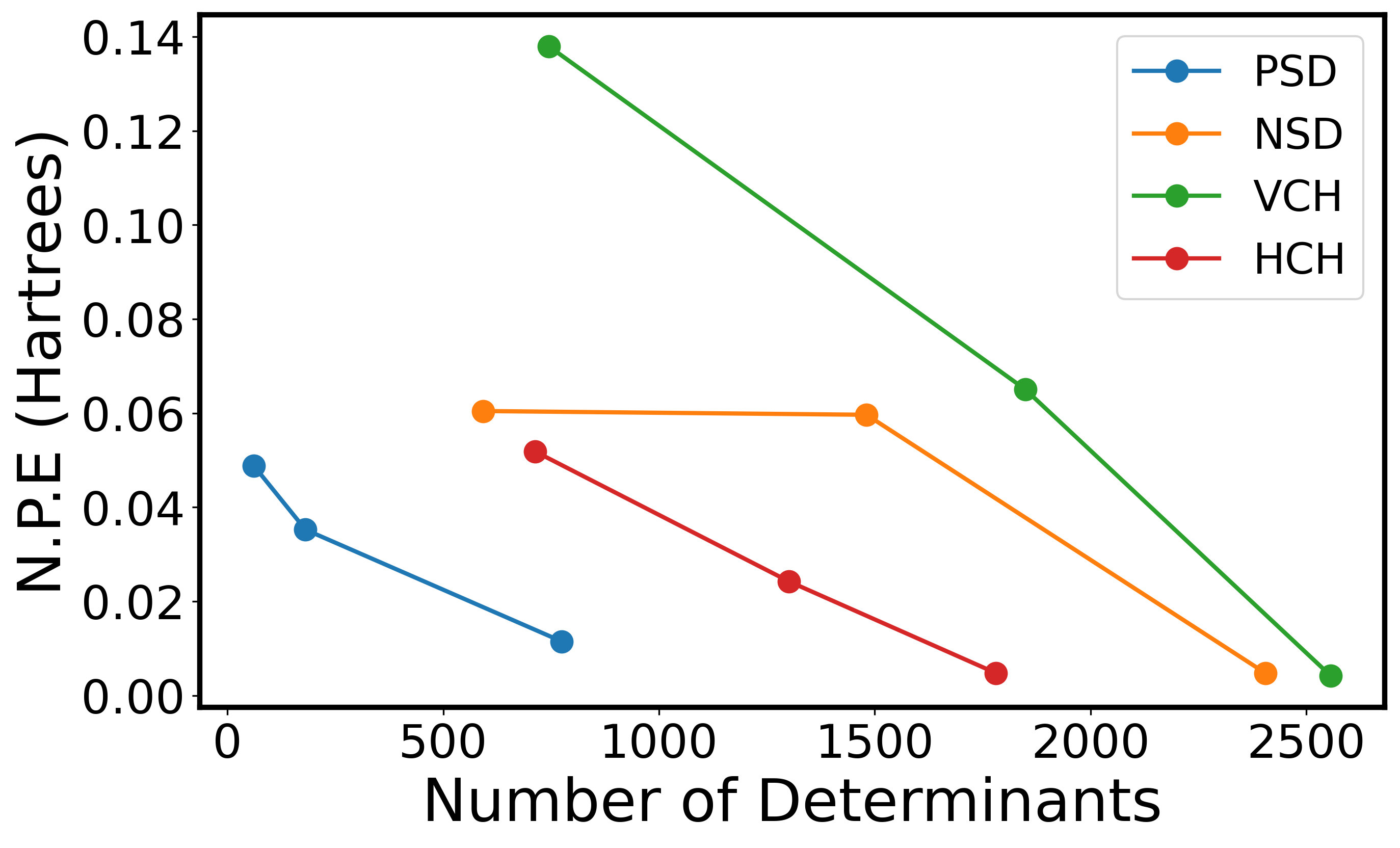} 
        \caption{}
        \label{fig:npe-a}
    \end{subfigure}
    \hfill
    \begin{subfigure}[b]{0.45\textwidth} 
        \centering
        \includegraphics[scale=0.335]{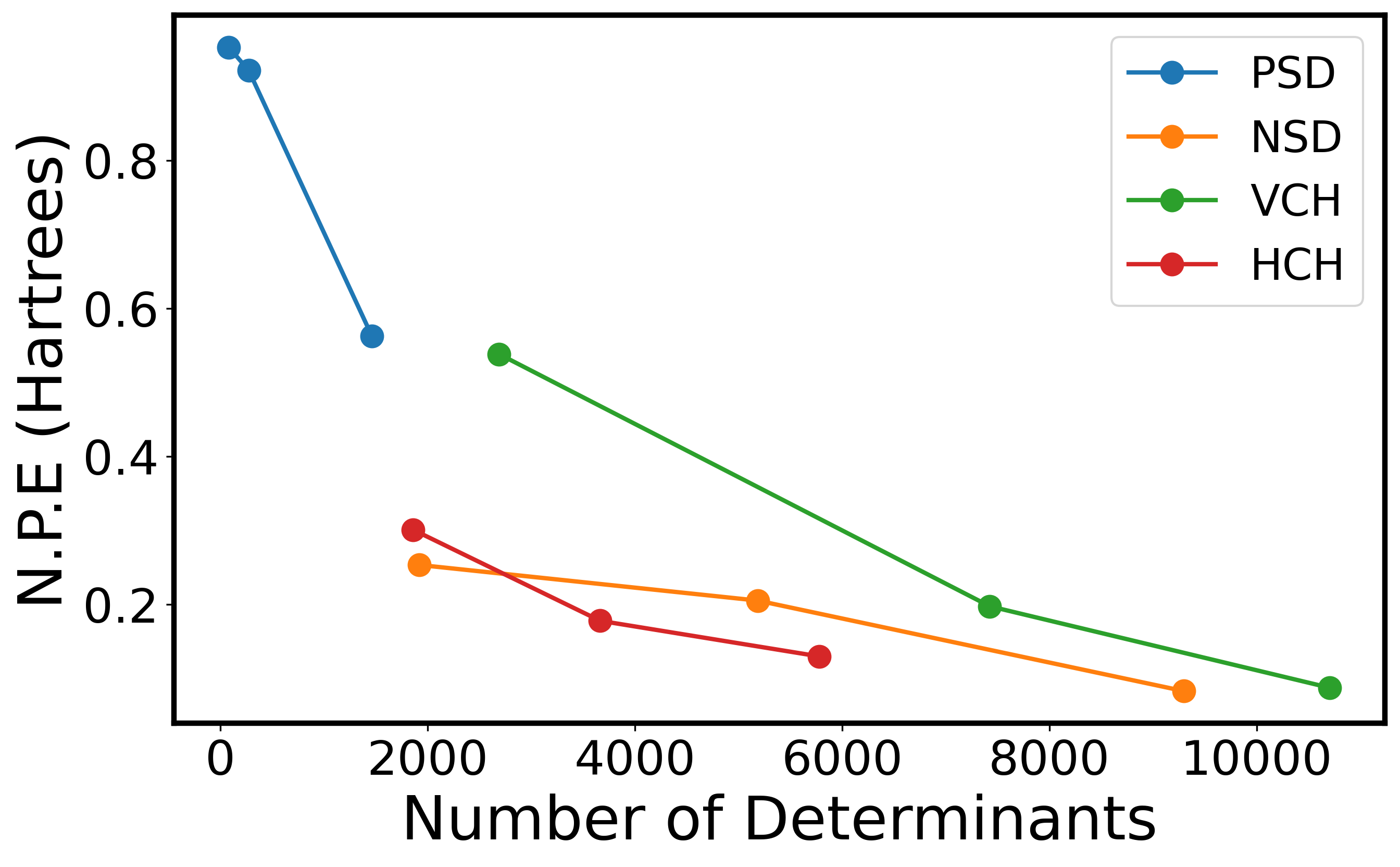} 
        \caption{}
        \label{fig:npe-b}
    \end{subfigure}
    \begin{subfigure}[b]{0.45\textwidth} 
        \centering
        \includegraphics[scale=0.335]{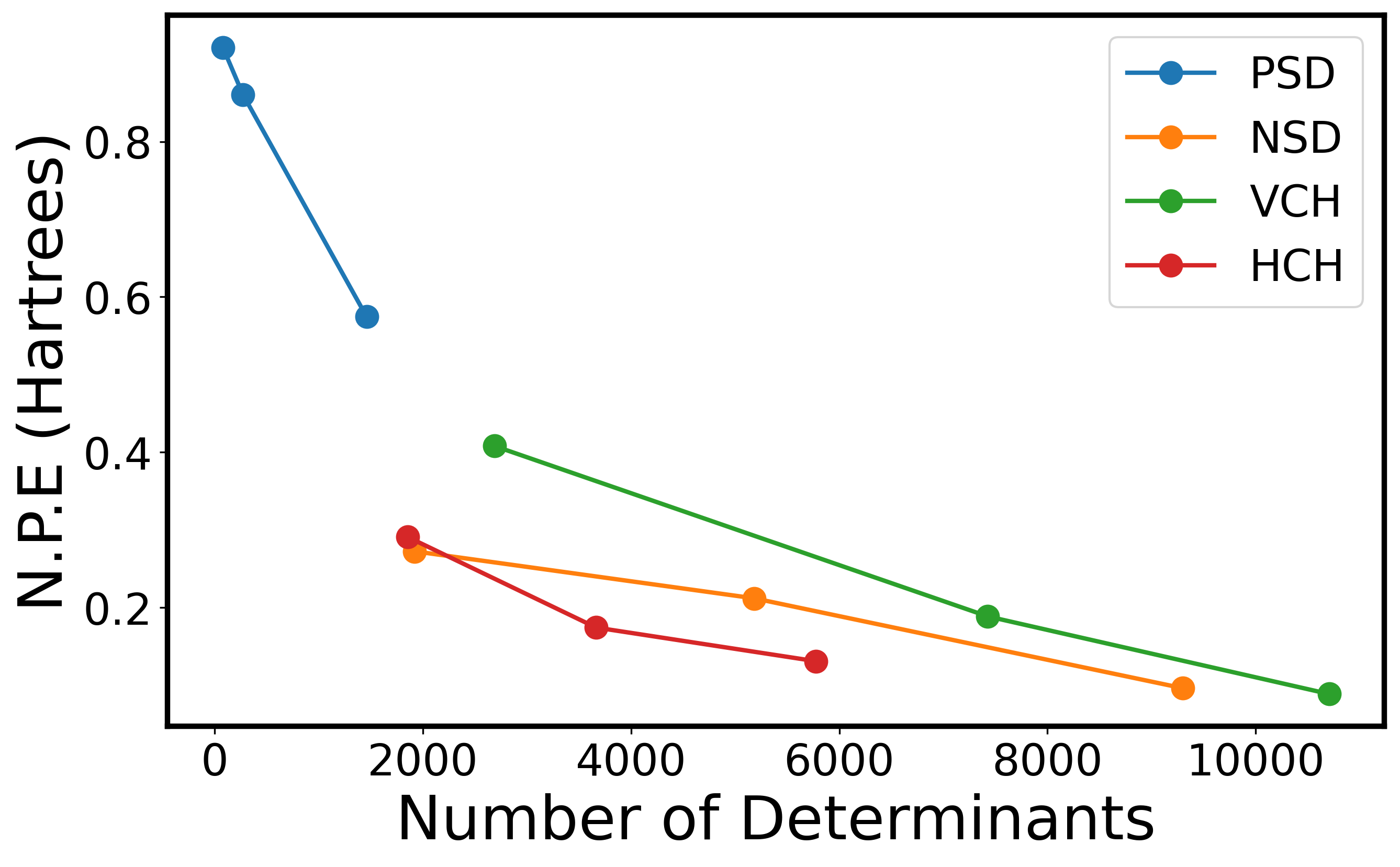} 
        \caption{}
        \label{fig:npe-c}
    \end{subfigure}

\caption{Non-parallelity errors plotted as a function of the number of determinants for the four partitioning schemes employed in the ehCI framework: PSD, NSD, VCH, and HCH. (a)Be\ce{H2} (b) Cubic \ce{H8} (c) Linear \ce{H8}}
\label{fig:npe}
\end{figure}

\autoref{fig:npe} presents the non‐parallelity error (NPE) as a function of the number of determinants retained for three model systems: (a) the C$_{2v}$ insertion of Be into \ce{H2}, (b) the cubic \ce{H8} cluster, and (c) the linear \ce{H8} chain.
The NPE\ is defined as the difference between the maximum and minimum point‐wise deviations of an approximate potential‐energy curve from the FCI reference,
\[
\mathrm{NPE} \;=\;\max_{R}\bigl[\Delta E(R)\bigr]\;-\;\min_{R}\bigl[\Delta E(R)\bigr],
\]
and serves as a measure of the uniformity with which the energy curve tracks the FCI surface across the entire reaction coordinate.

For \ce{BeH2} (\autoref{fig:npe-a}), all four partitioning schemes ultimately converge to zero NPE as the configuration space approaches the FCI limit.
Among them, the PSD and HCH partitions achieve the smallest NPE values at any given number of determinants, indicating that these schemes reproduce the shape of the FCI potential energy curve most effectively when working with a relatively compact determinant space.
In contrast, the NSD scheme shows somewhat larger deviations (with an NPE of approximately 0.06~Hartree at around 1000 determinants), while the VCH partition exhibits the largest NPE (approximately 0.14~Hartree at a similar rank).
This pattern reflects each partition’s ability to incorporate critical seniority-zero configurations that are essential for capturing static correlation.
These trends make clear that agreement with the FCI shape does not depend solely on the total number of determinants but rather on the inclusion of the determinant sectors that contribute most significantly to both static and dynamic correlation.

When the system size increases to eight electrons in the H$_8$ cube (\autoref{fig:npe-b}), the total determinant space expands combinatorially, and the challenge of maintaining parallelity becomes more pronounced.
In this regime, NSD and HCH continue to provide the lowest NPEs across truncation levels (e.g., approximately 0.25~Hartree for around 2000 determinants), while the VCH and especially the PSD partition show larger deviations (around 0.55~Hartree for the same rank).
Although all schemes show a general reduction in NPE as more determinants are added, their relative performance remains consistent, emphasizing that it is the selection of determinants, particularly those balancing excitation rank with seniority, that primarily governs shape accuracy in larger, strongly correlated systems.
The linear H$_8$ dissociation (\autoref{fig:npe-c}) yields similar observations: NSD and HCH partitions deliver consistently lower NPEs over a wide range of determinant counts, while VCH is intermediate and PSD performs poorest at lower truncation levels.

Taken together, these results reinforce that carefully balancing excitation degree and seniority within the hierarchy is essential for achieving low NPE errors.
Schemes that admit seniority-zero, seniority-two, or seniority-four configurations at lower hierarchy levels preserve the correct qualitative features of the potential energy curve with far fewer determinants than schemes that delay these sectors until later stages.
Including the determinants that strongly couple to the reference and capture near-degeneracies and static correlation effects is more effective than merely enlarging the configuration space without regard to its composition.
It highlights the practical advantage of partitioning schemes that are designed to admit these critical blocks early in the hierarchy, enabling compact yet accurate CI expansions that reproduce the FCI shape more reliably.

\subsection{Effect of Ad Hoc Inclusion of Off-Hierarchy Determinants}
Beyond the standard linear partitioning defined by $(\alpha_1, \alpha_2)$, the inherent flexibility of the extended hierarchy CI (ehCI) framework, together with the modular design of the FANCI and Fanpy packages, enables straightforward exploration of additional wavefunctions by selectively expanding the determinant subspace with configurations that formally lie outside the nominal hierarchy.
For example, in the VCH scheme with $h = 0$, the base subspace excludes certain key single and double excitations, specifically, sectors such as ($e = 1$, $s = 2$), ($e = 2$, $s = 0$), ($e = 2$, $s = 2$), and ($e = 2$, $s = 4$) which would otherwise contribute significant static or dynamic correlation.
The figures presented below illustrate the impact of selectively including specific off-hierarchy determinant blocks, characterized by excitation degree ($e$) and seniority ($s$), in different ehCI partitioning schemes.
For each scheme, the VCH, HCH, and NSD, the base hierarchy at a given $h$ value defines a systematic truncation of the Hilbert space. 
However, additional determinant sectors that do not formally belong to the nominal hierarchy are introduced to examine how their targeted inclusion affects the total energy and potential energy curve shape relative to the FCI benchmark.
Each curve shows the HF reference, the base hierarchy result, and the hierarchy result with specific ($e$,$s$) blocks added, allowing direct comparison of how these sectors contribute to recovering static and dynamic correlation effects across the reaction coordinate.
This comparative analysis highlights the interplay between excitation degree, seniority, and determinant coupling pathways in improving wavefunction flexibility and lowering the variational energy. 

\begin{figure}[H] 
    \centering
    \begin{subfigure}[b]{0.45\textwidth} 
        \centering
        \includegraphics[scale=0.335]{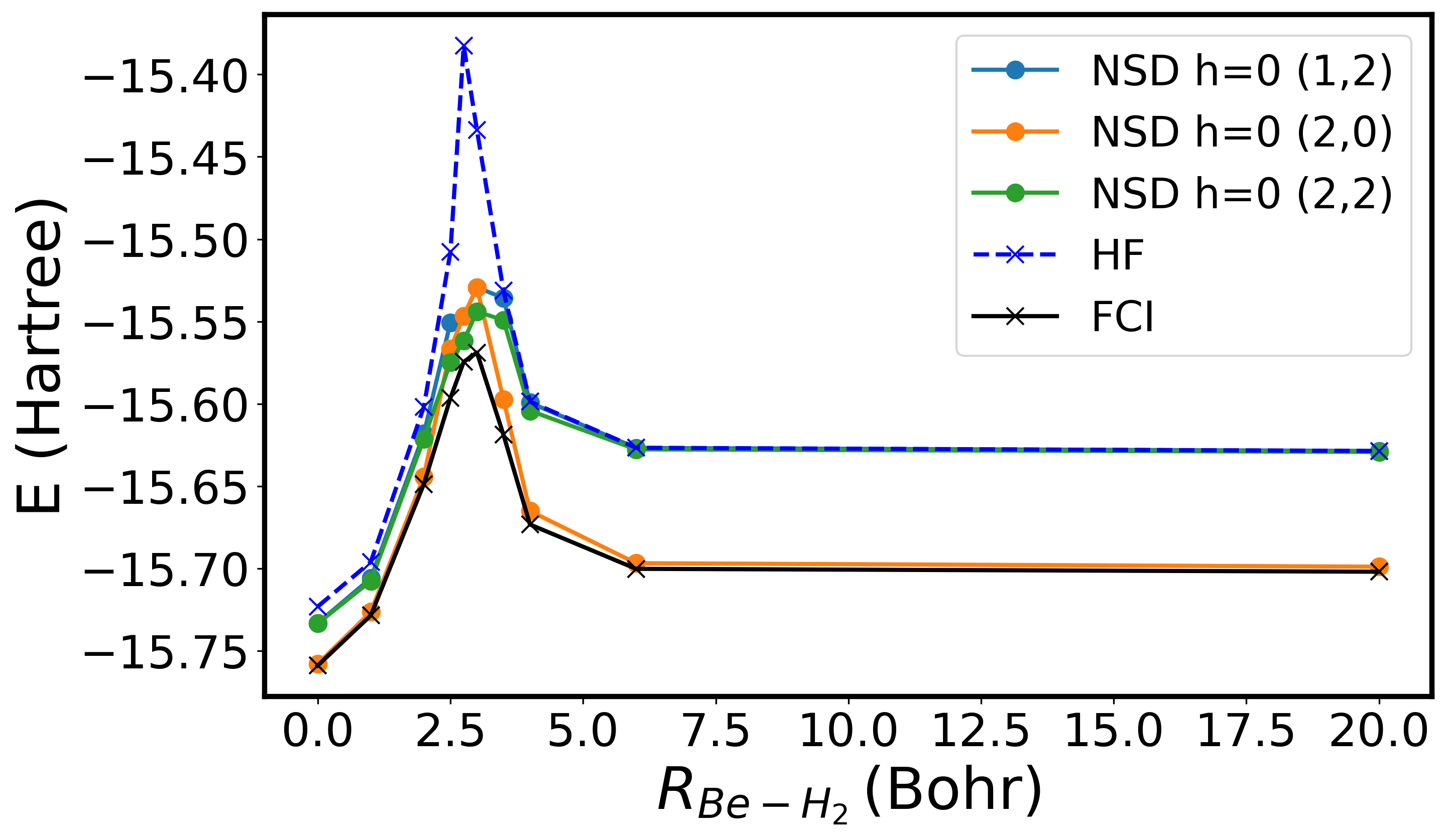} 
        \label{fig:image26}
    \end{subfigure}
    \hfill
    \begin{subfigure}[b]{0.45\textwidth} 
        \centering
        \includegraphics[scale=0.335]{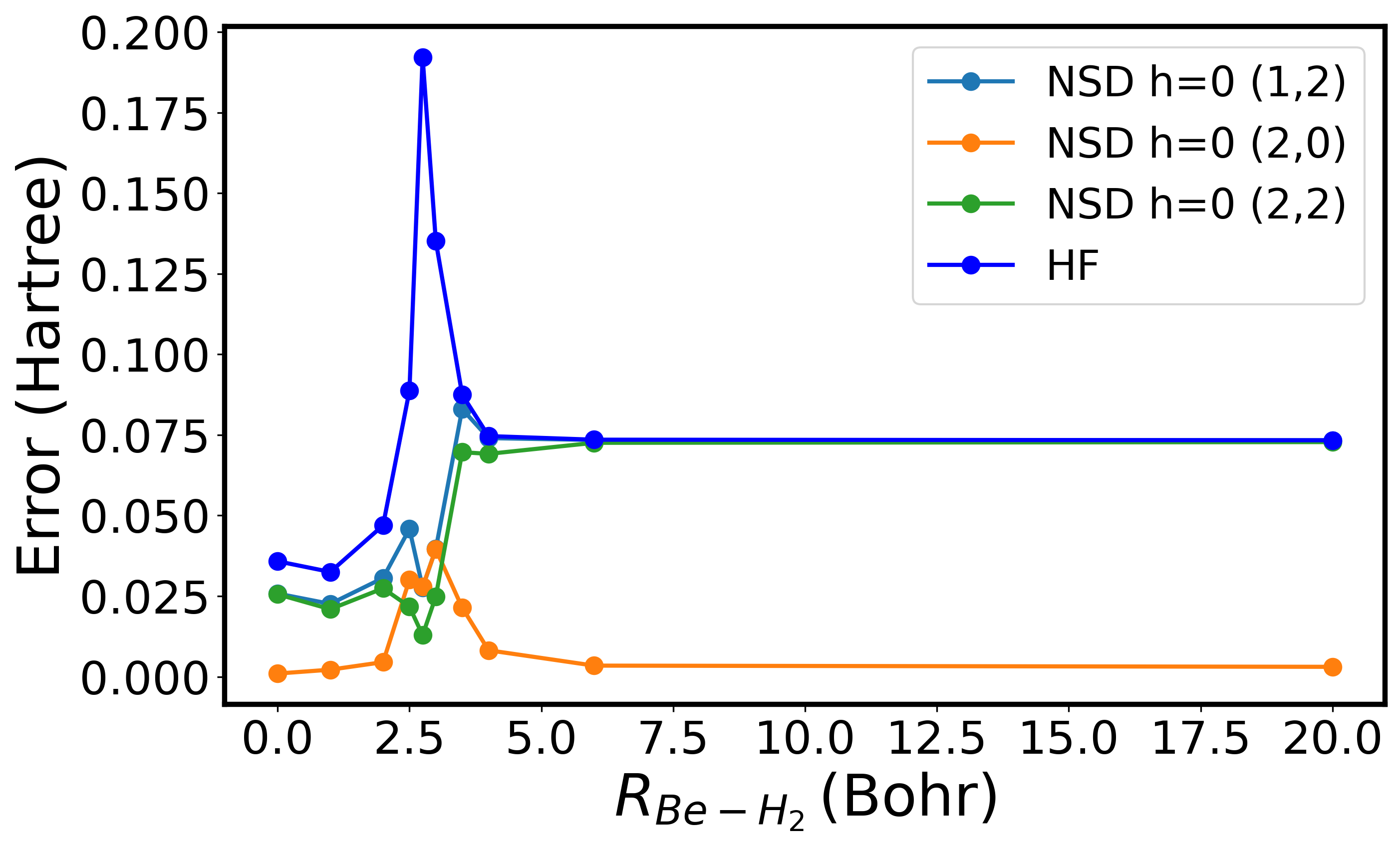} 
        \label{fig:image27}
    \end{subfigure}
    \begin{subfigure}[b]{0.45\textwidth} 
        \centering
        \includegraphics[scale=0.335]{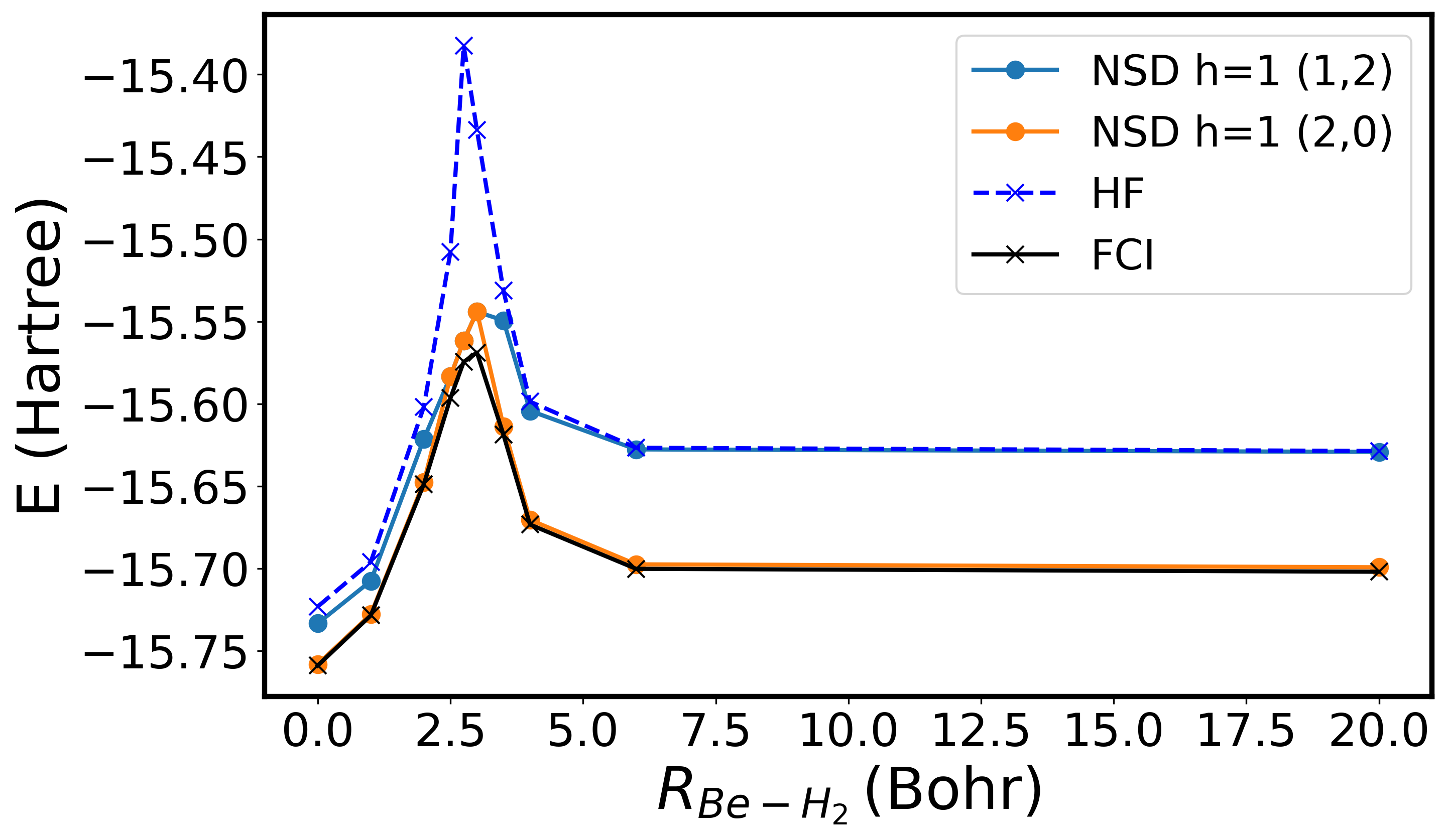} 
        \label{fig:image28}
    \end{subfigure}
    \hfill
    \begin{subfigure}[b]{0.45\textwidth} 
        \centering
        \includegraphics[scale=0.335]{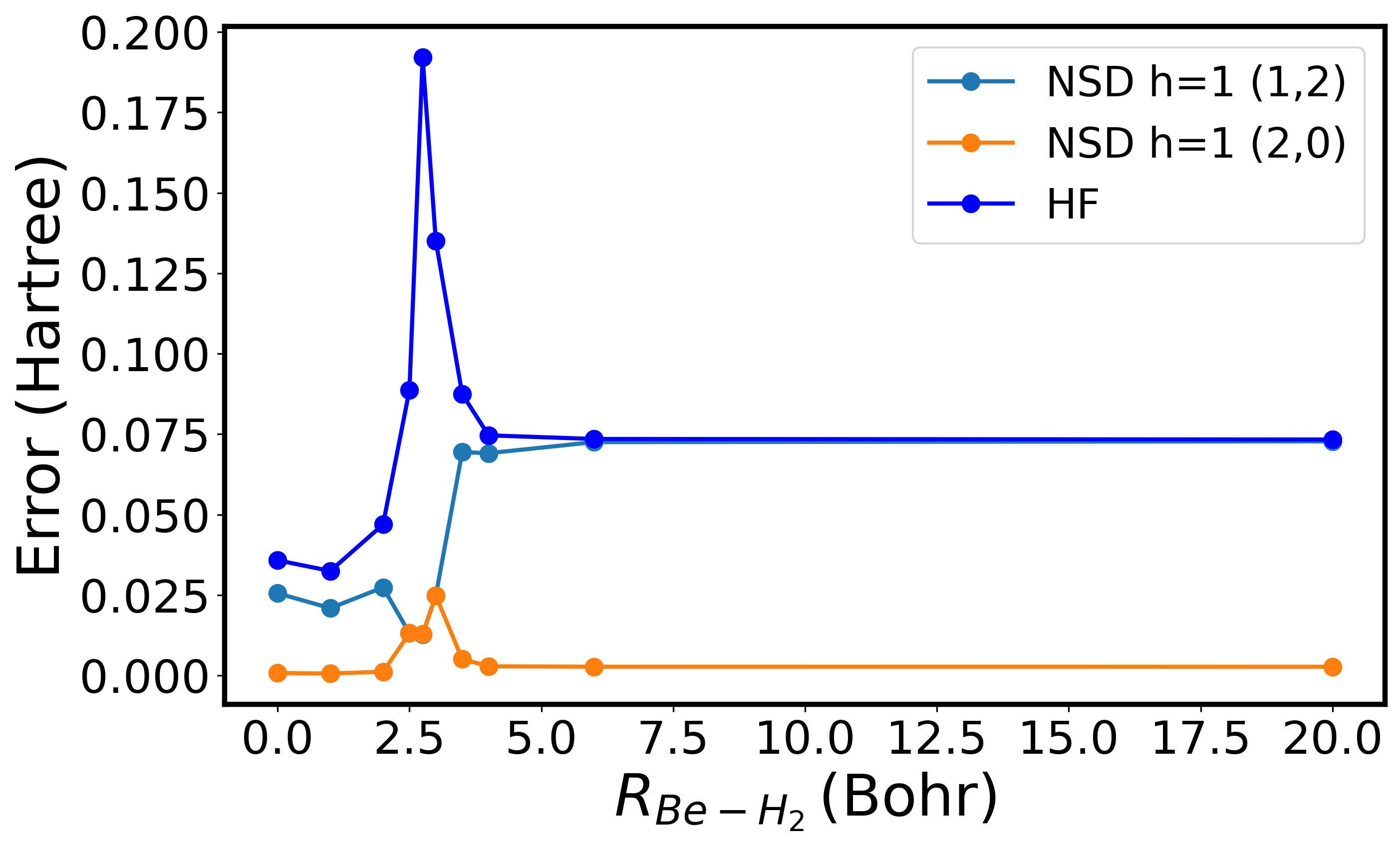} 
        \label{fig:image29}
    \end{subfigure}
    \begin{subfigure}[b]{0.45\textwidth} 
        \centering
        \includegraphics[scale=0.335]{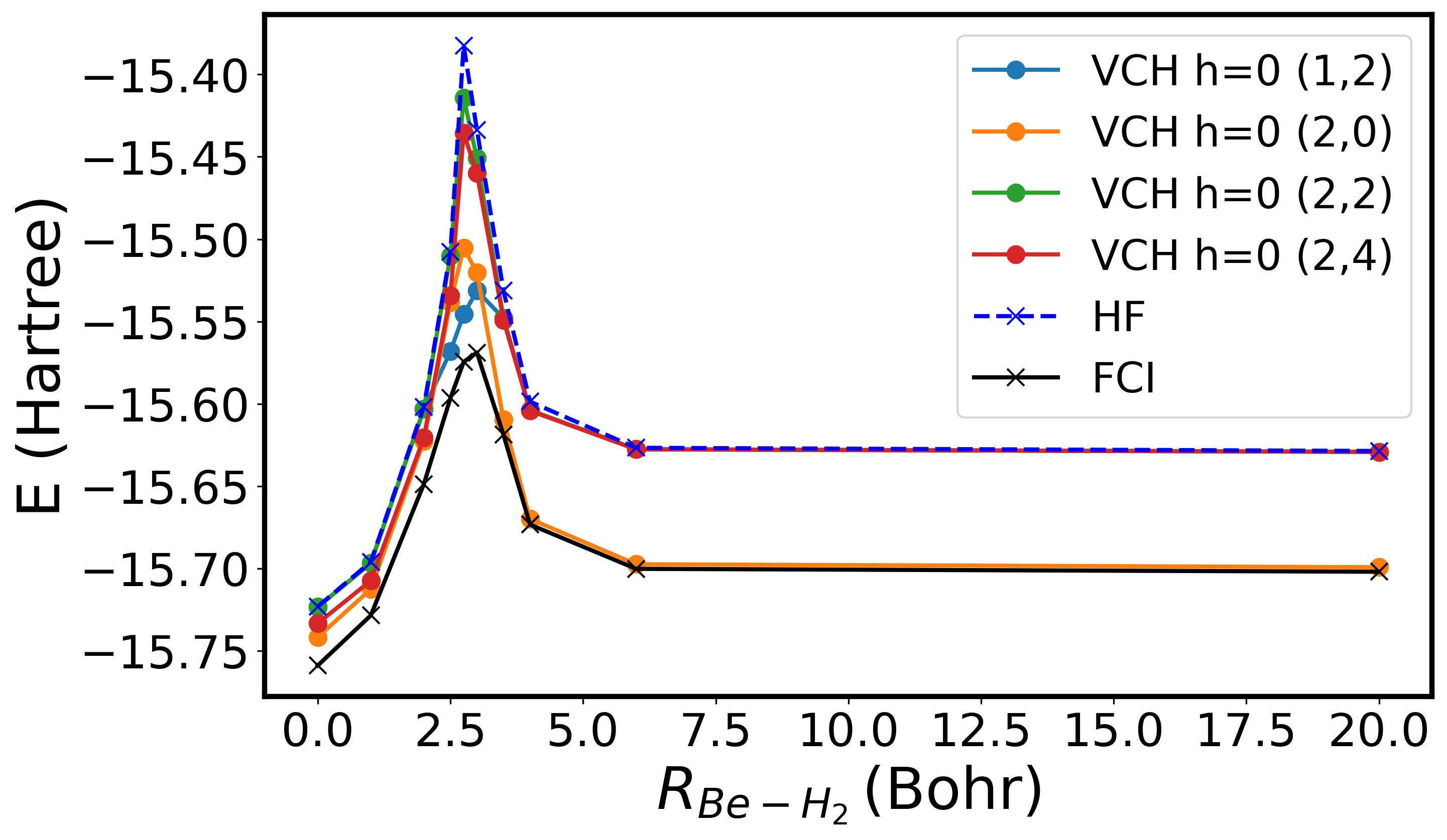} 
        \label{fig:image30}
    \end{subfigure}
    \hfill
    \begin{subfigure}[b]{0.45\textwidth} 
        \centering
        \includegraphics[scale=0.335]{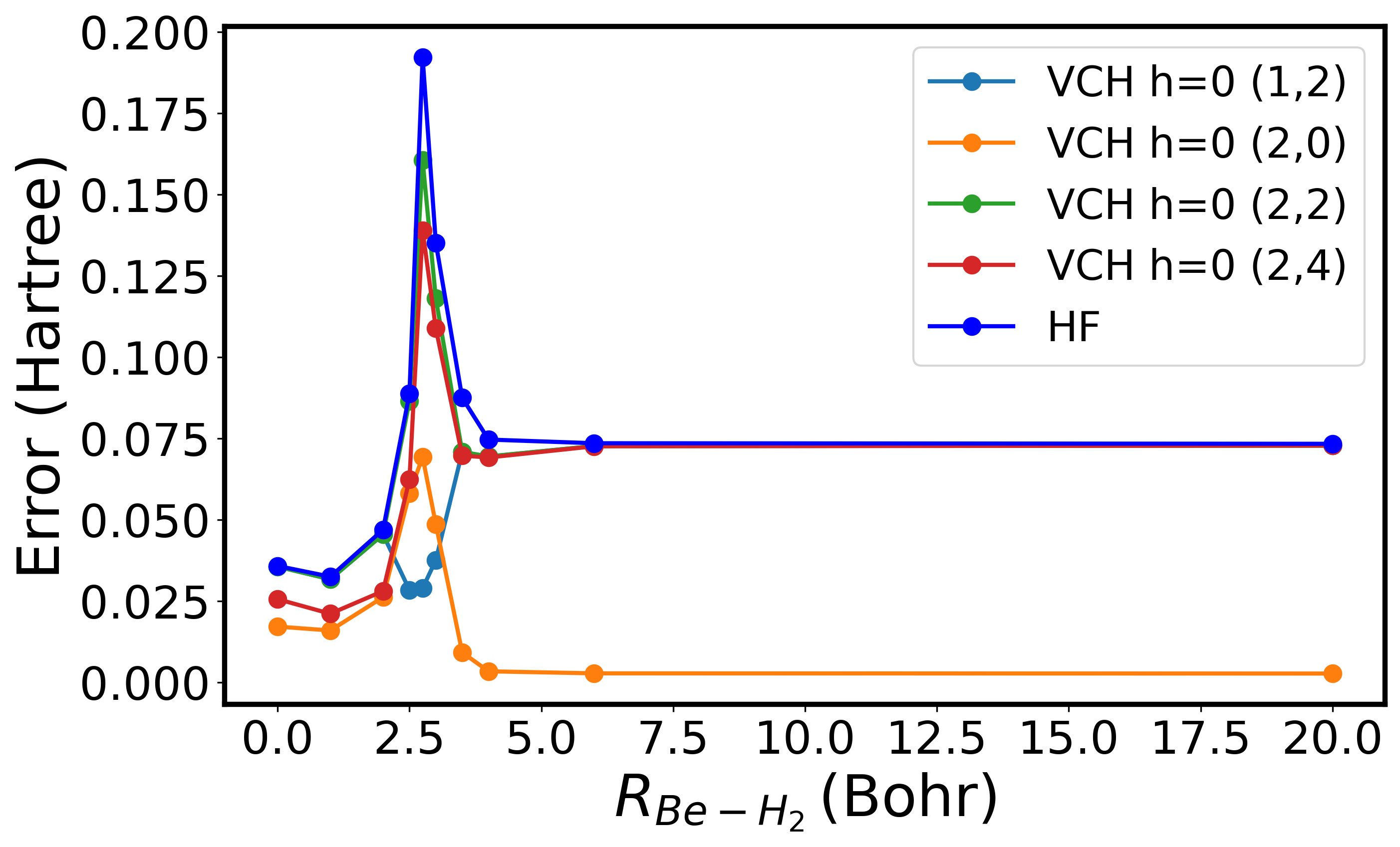} 
        \label{fig:image31}
    \end{subfigure}
    \begin{subfigure}[b]{0.45\textwidth} 
        \centering
        \includegraphics[scale=0.335]{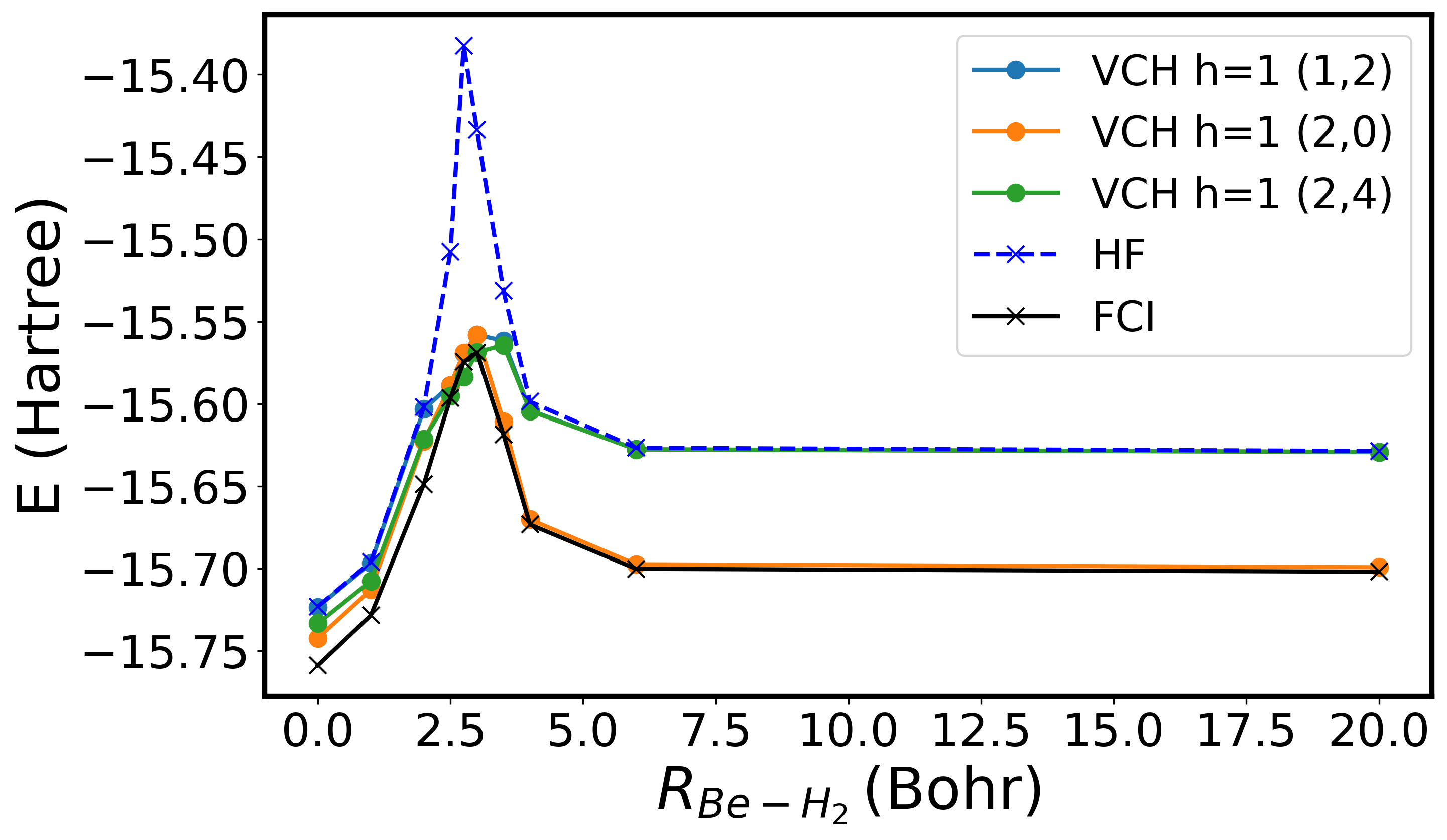} 
        \label{fig:image32}
    \end{subfigure}
    \hfill
    \begin{subfigure}[b]{0.45\textwidth} 
        \centering
        \includegraphics[scale=0.335]{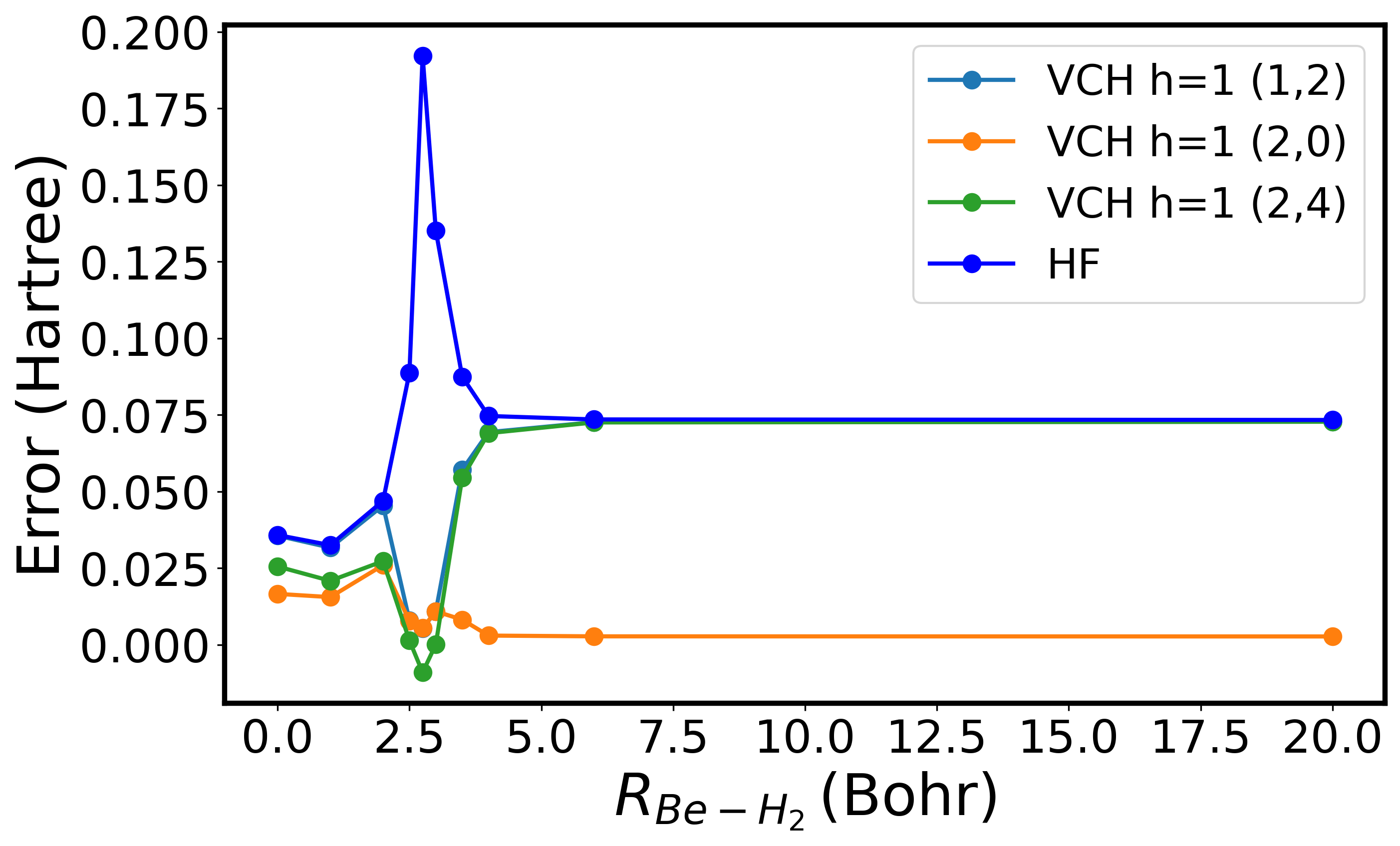} 
        \label{fig:image33}
    \end{subfigure}
    \label{ad_hoc_addition}
\end{figure}

\begin{figure}[H] 
    \centering
    \hfill
    \begin{subfigure}[b]{0.45\textwidth} 
        \centering
        \includegraphics[scale=0.335]{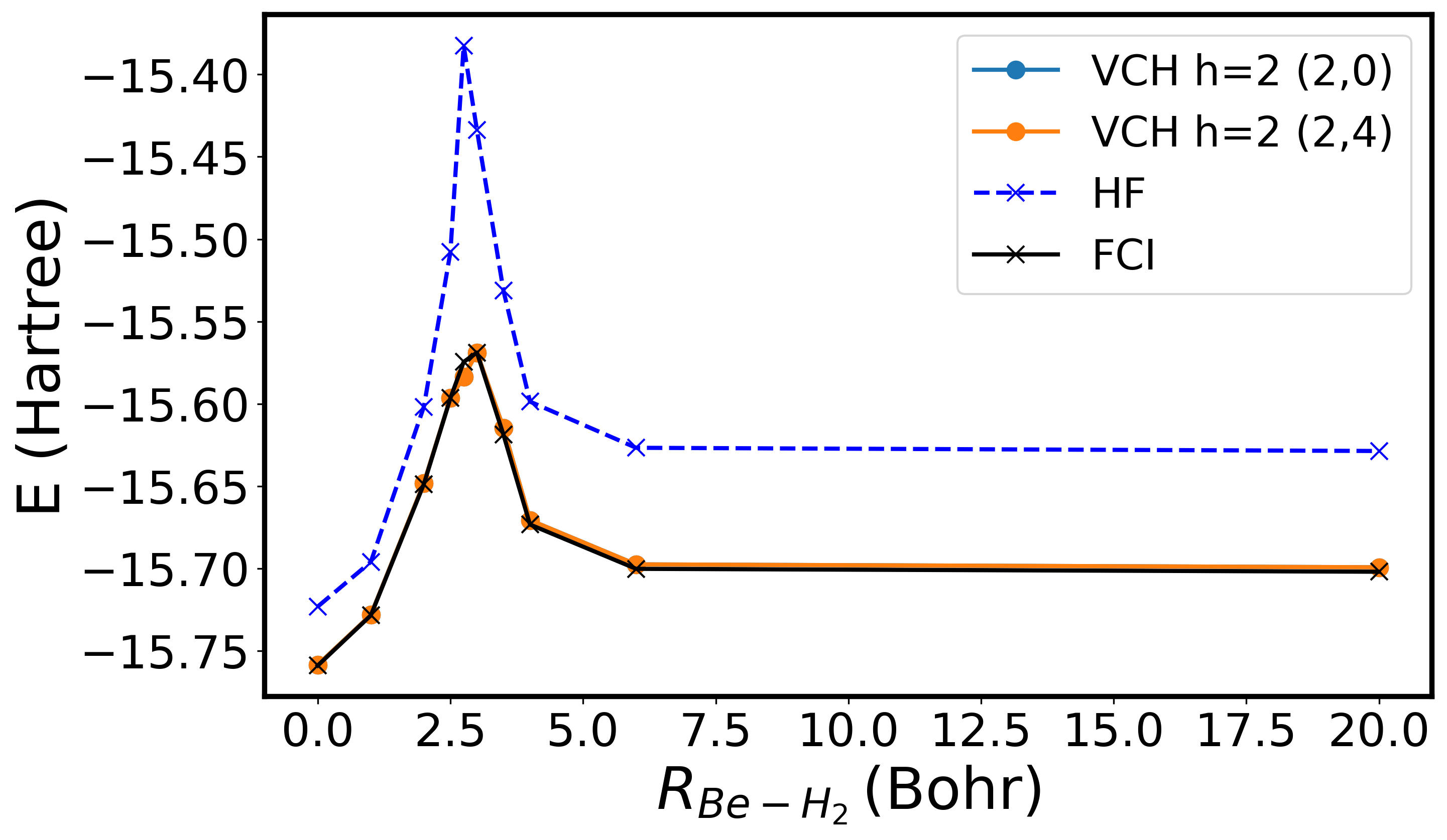} 
        \label{fig:image34}
    \end{subfigure}
    \hfill
    \begin{subfigure}[b]{0.45\textwidth} 
        \centering
        \includegraphics[scale=0.335]{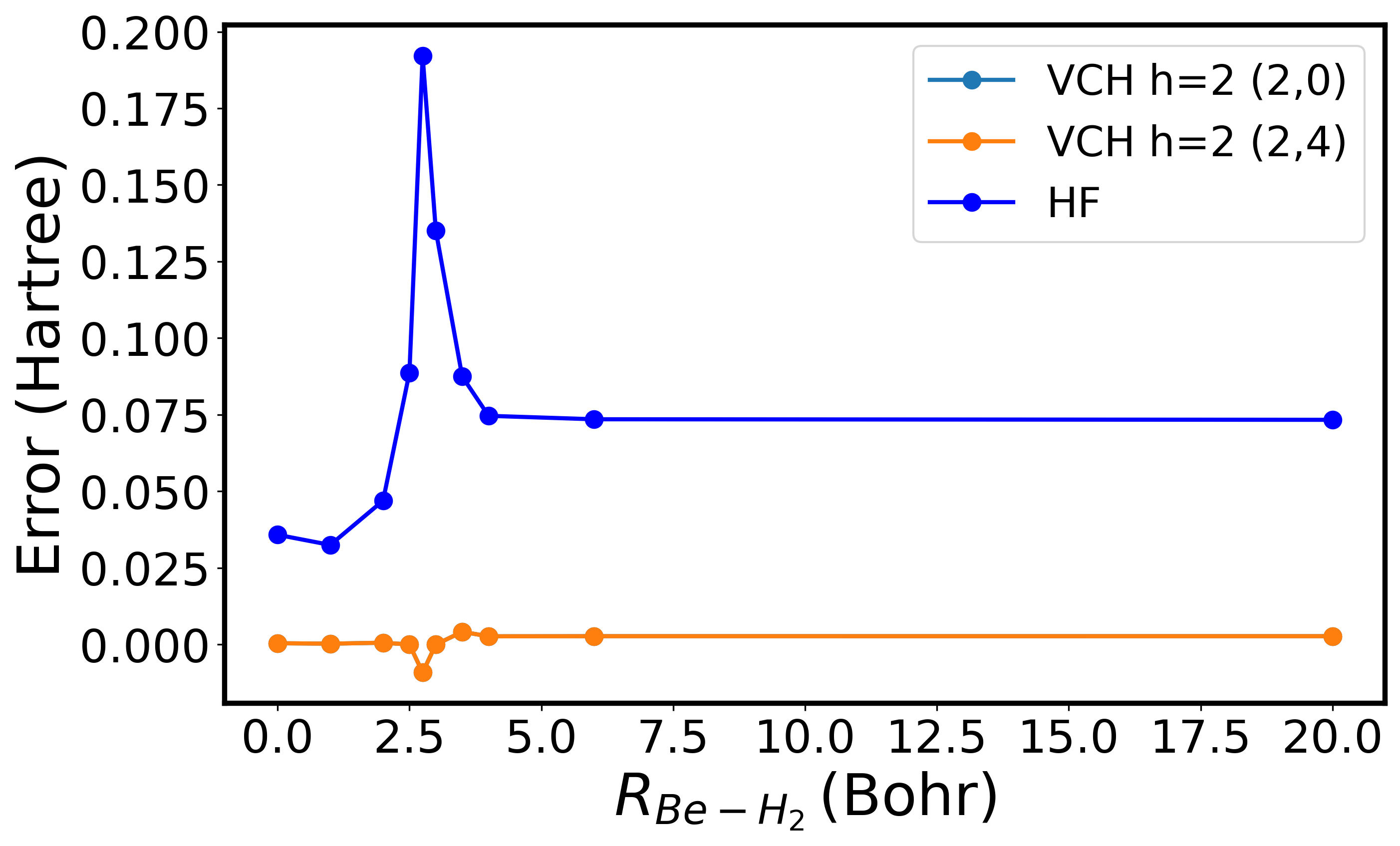} 
        \label{fig:image35}
    \end{subfigure}
    \hfill
    \begin{subfigure}[b]{0.45\textwidth} 
        \centering
        \includegraphics[scale=0.335]{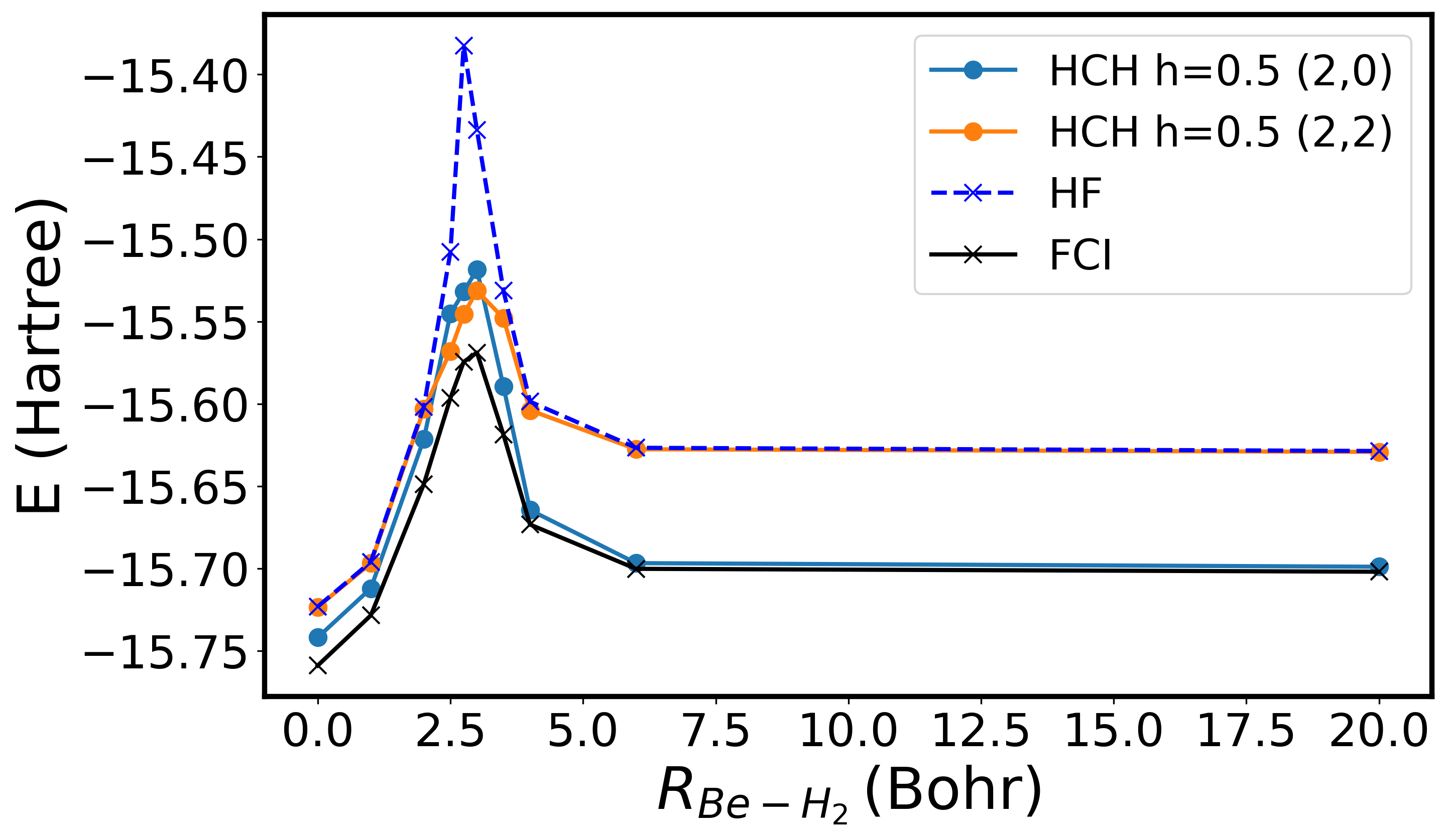} 
        \label{fig:image36}
    \end{subfigure}
    \hfill
    \begin{subfigure}[b]{0.45\textwidth} 
        \centering
        \includegraphics[scale=0.335]{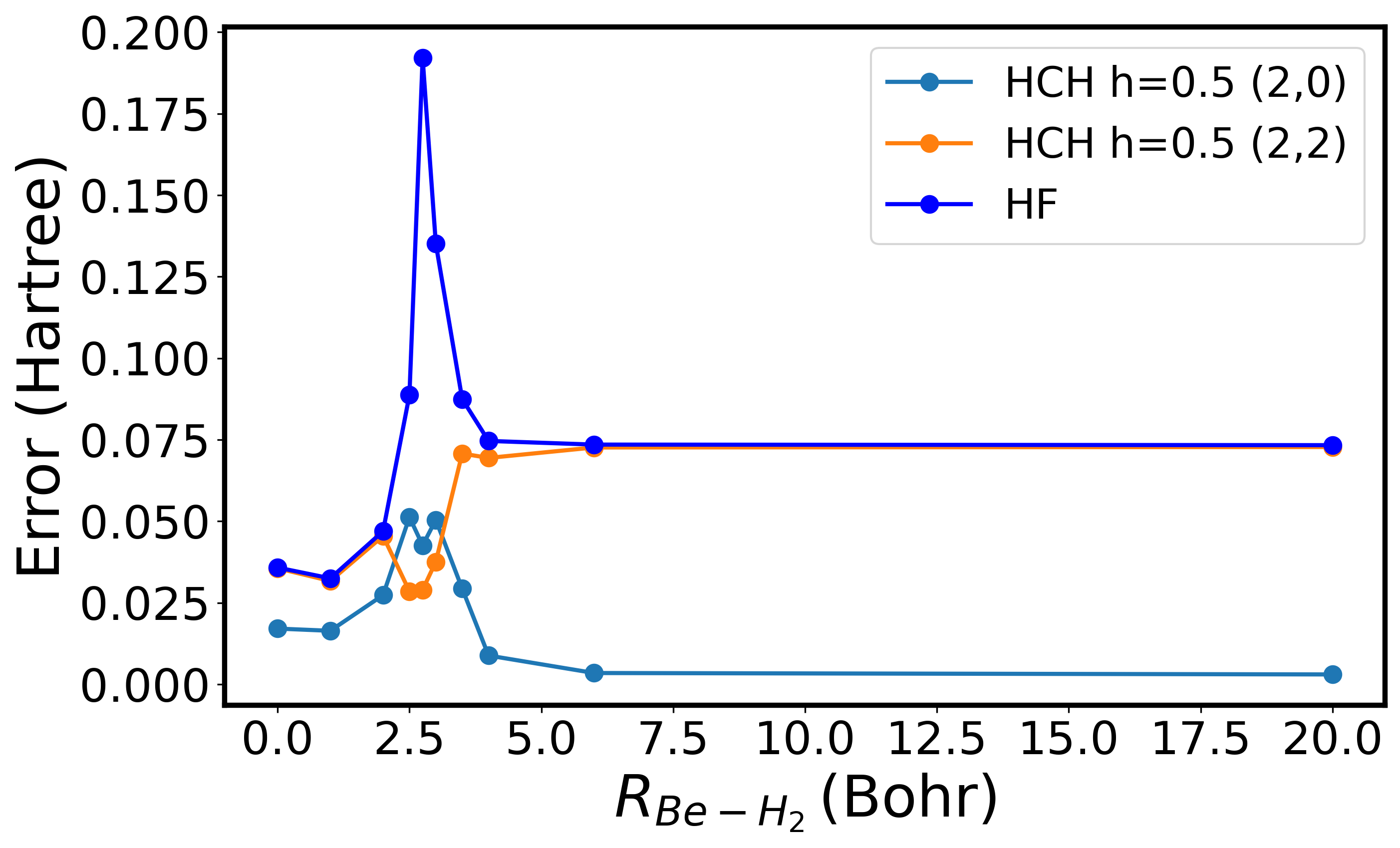} 
        \label{fig:image37}
    \end{subfigure}
    \hfill
    \begin{subfigure}[b]{0.45\textwidth} 
        \centering
        \includegraphics[scale=0.335]{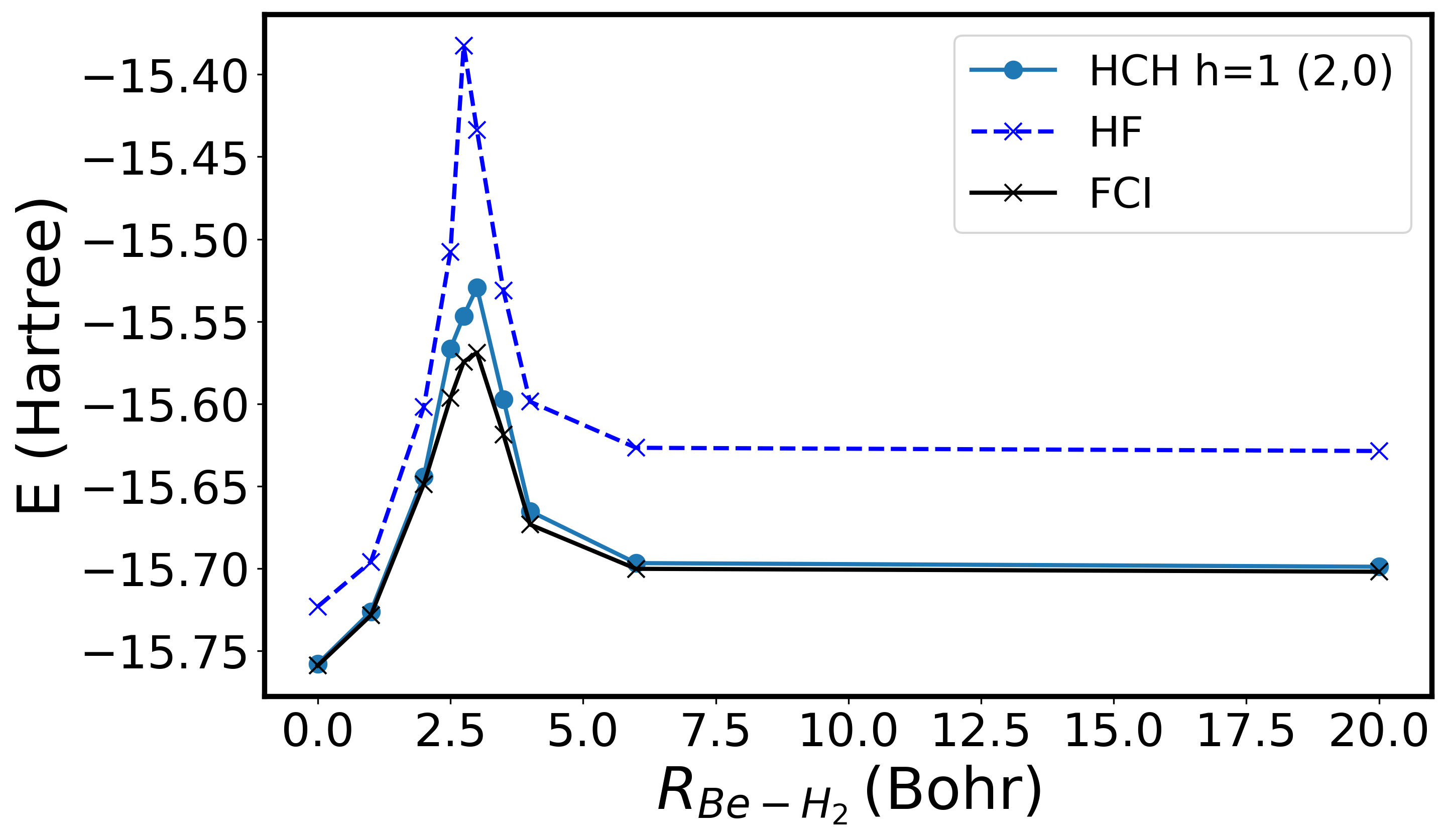} 
        \label{fig:image38}
    \end{subfigure}
    \hfill
    \begin{subfigure}[b]{0.45\textwidth} 
        \centering
        \includegraphics[scale=0.335]{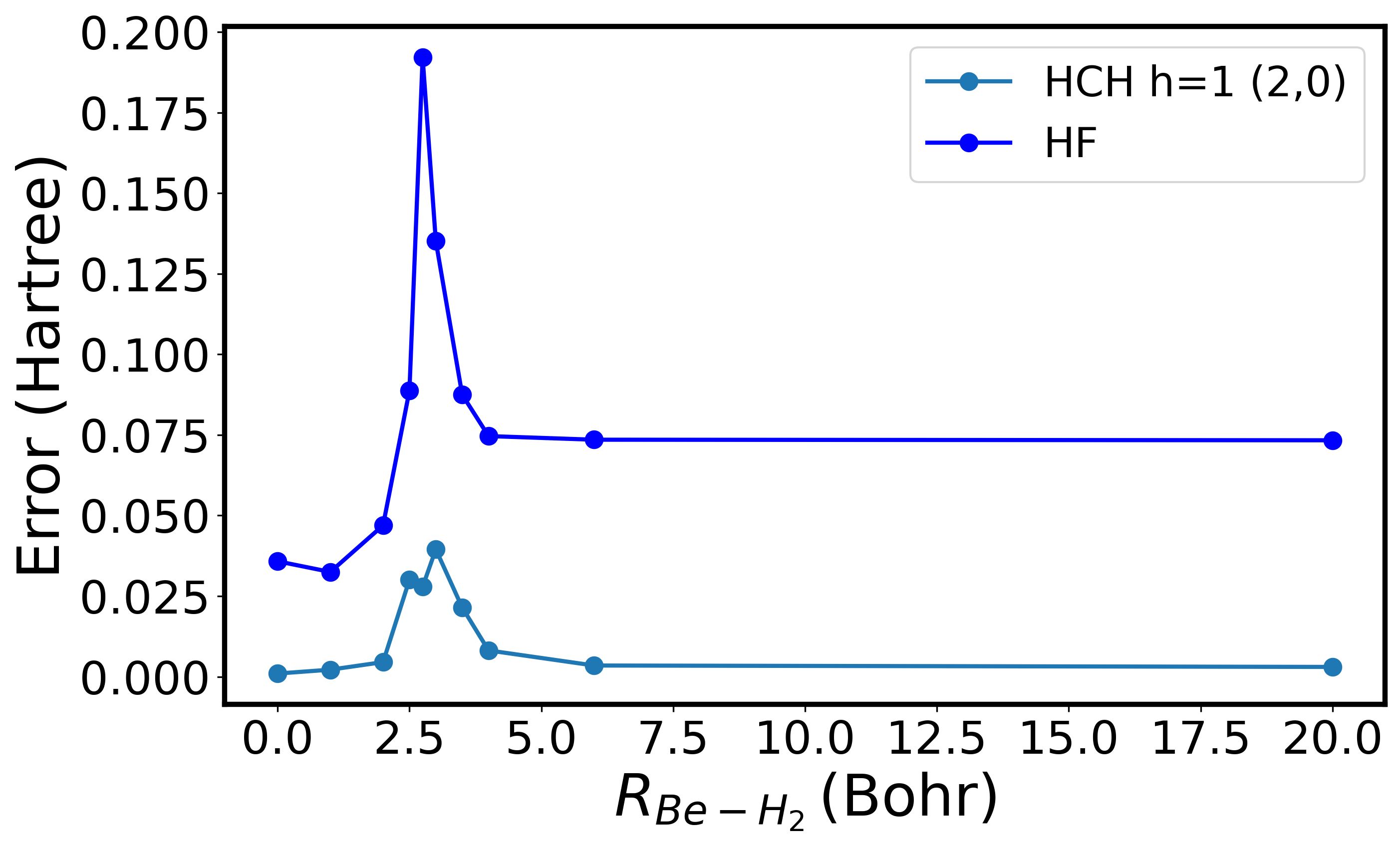} 
        \label{fig:image39}
    \end{subfigure}
    \hfill
    \begin{subfigure}[b]{0.45\textwidth} 
        \centering
        \includegraphics[scale=0.335]{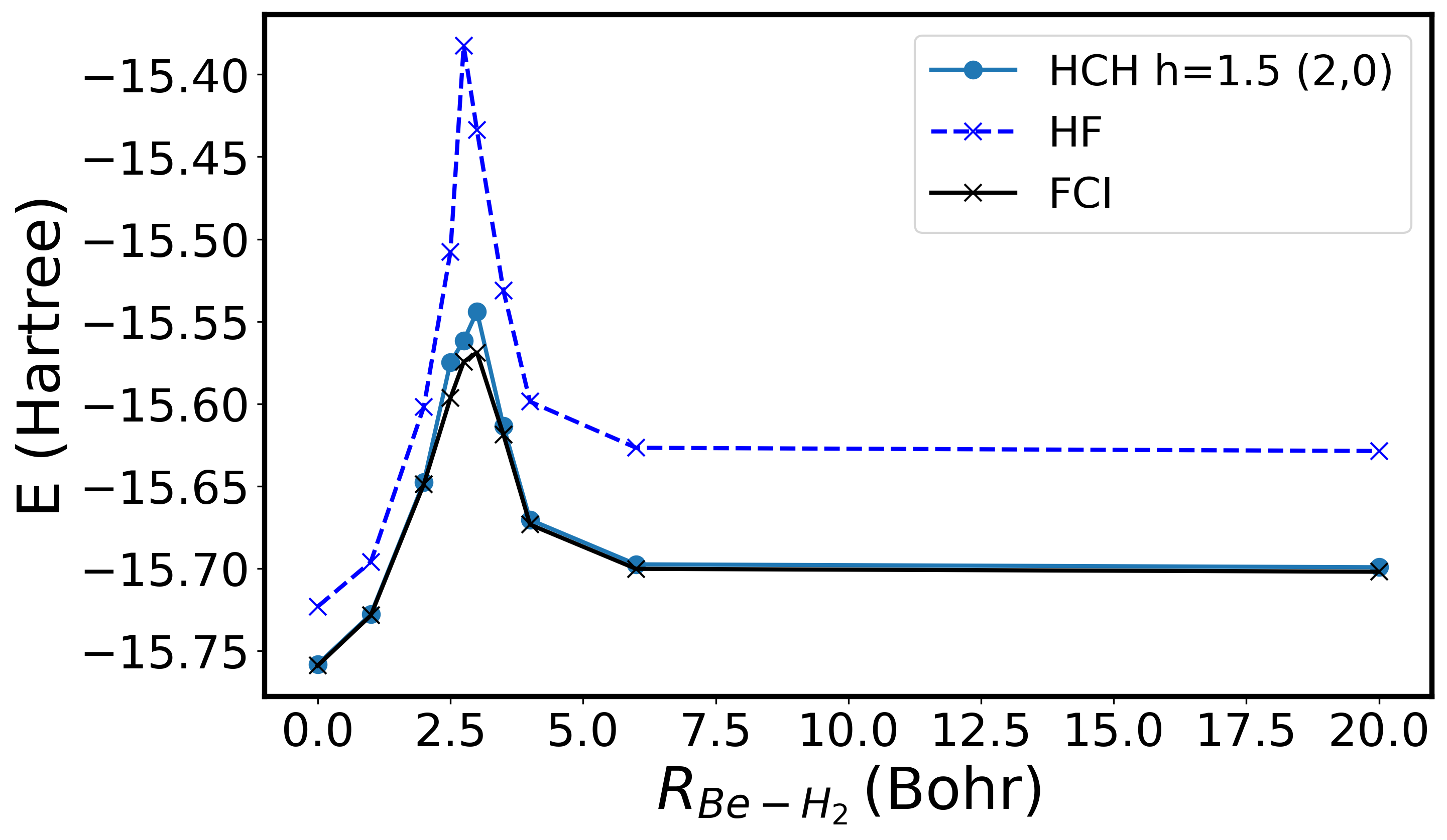} 
        \label{fig:image40}
    \end{subfigure}
    \hfill
    \begin{subfigure}[b]{0.45\textwidth} 
        \centering
        \includegraphics[scale=0.335]{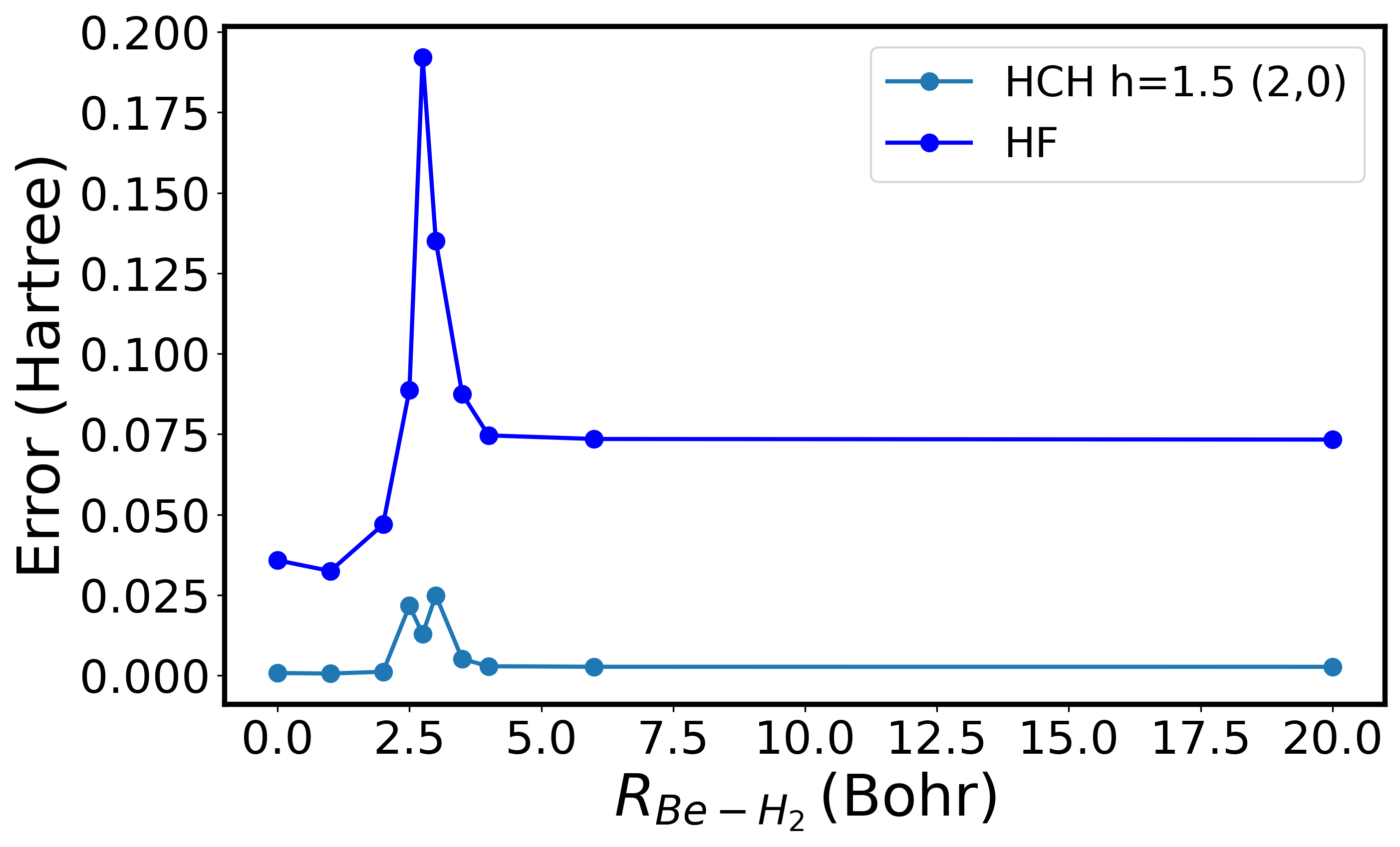} 
        \label{fig:image41}
    \end{subfigure}
    \caption{
Results for the VCH, HCH, and NSD scheme at a given hierarchy level $h$, with additional determinant sectors characterized by excitation degree ($e$) and seniority ($s$) included beyond the nominal hierarchy.
The curves compare the Hartree-Fock (HF) reference, the base hierarchy result, and the hierarchy result with specific ($e$,$s$) blocks added.
}
    \label{ad-hoc}
\end{figure}
As mentioned above, in the VCH scheme, the base space naturally excludes low-excitation, seniority-zero double excitations (2,0), which are essential for capturing static correlation through pairwise promotions that strongly couple to the HF reference.
Adding these blocks significantly lowers the total energy and improves the quality of the potential energy curve relative to FCI, especially in strongly correlated regions.
Including higher-seniority doubles such as (2,2) or (2,4) further refines the dynamic correlation, but with a smaller effect than the dominant (2,0) sector.

From a wavefunction perspective, this effect can be rationalized by considering the coupling structure in the CI Hamiltonian.
Static correlation often requires the inclusion of determinants that maintain seniority zero or two, especially in multi-reference regimes such as bond-breaking or insertion pathways.
The seniority-zero block (e.g., (2, 0)) includes pairwise double excitations that preserve spin pairing and can strongly couple to the reference determinant via the two-electron interaction operator. 
Likewise, seniority-two blocks like  (1, 2), (2, 2) capture configurations with partially broken pairs, which are essential for describing near-degeneracies when multiple determinants become quasi-degenerate with the reference.

A similar trend can be found in the HCH hierarchy, which initially emphasizes low excitation degree but may postpone critical low-seniority pairs.
In the base HCH, the absence of (2, 0) block means that static correlation is poorly described.
When (2, 0) is added off-hierarchy, the wavefunction gains direct coupling pathways to the reference determinant, yielding substantial energy lowering and better agreement with FCI.
Adding (2, 2) alongside or after (2,0) further enhances the accuracy by incorporating one broken pair configuration that captures near-degeneracy and asymmetric correlation effects, but the static correlation channel from (2, 0) remains the most impactful.

For the NSD scheme, which balances excitation and seniority in diagonal slices, adding off-hierarchy blocks shows how each sector complements the base space.
NSD at low $h$ already mixes in singles and some doubles but may miss key static pair excitations if (2, 0) is delayed.
Including (2, 0) immediately recovers the dominant static correlation missing from the HF reference, while adding (2, 2) helps adjust the fine details of dynamic correlation.
These selective additions highlight that different ($e$,$s$) blocks interact variationally through the CI Hamiltonian: seniority-zero doubles strongly mix with the reference, while higher-seniority or more complex excitations connect through intermediate configurations, providing incremental improvements.

Overall, these results underscore that rigid partitioning, while systematic, may fail to capture system-specific correlation efficiently at low hierarchy levels.
Selective inclusion of off-hierarchy sectors, particularly those with low excitation degree and low seniority, helps recover the missing static and dynamic components, leading to smoother potential energy surfaces with reduced errors.
This suggests that practical implementations of extended hierarchy CI (ehCI) could benefit from the adaptive promotion of determinant blocks based on coupling strengths or perturbative estimates, bridging the gap between strict hierarchy schemes and more flexible selected CI approaches.

\section*{Conclusions}

This work introduced the extended hierarchy configuration interaction (ehCI) framework, which combines excitation degree ($e$) and seniority number ($s$) into a single adjustable hierarchy parameter, $h = \alpha_1 e + \alpha_2 s$. By tuning the weights $(\alpha_1, \alpha_2)$, this approach unifies the strengths of excitation-based and seniority-based truncations while providing systematic control over the selection of determinants that recover both static and dynamic correlations. The flexibility of the hierarchy enables compact and accurate expansions that adapt to the correlation characteristics of different electronic systems.

Benchmark results for the \ce{BeH2} insertion pathway and the cubic and linear \ce{H8} dissociation curves demonstrate the effectiveness of this framework. Increasing the hierarchy cutoff ($h_\mathrm{max}$) produced smooth, monotonic convergence to the full configuration interaction (FCI) limit. Strategies that emphasize low-seniority pair excitations, such as the Negative Slope Diagonal and Horizontal Chess Horse schemes, achieved lower absolute errors and non-parallelity errors (NPEs), especially in regions of strong static correlation. In contrast, partitions that delay the inclusion of critical determinant classes required significantly larger expansions to reach comparable accuracy.

These observations highlight that the careful choice of determinant classes is more decisive than unrestricted expansion of the configuration space. Tests that selectively added off-hierarchy blocks further support the conclusion that combining the systematic hierarchy with targeted inclusion of key determinants can deliver a better balance of accuracy and efficiency. Overall, treating excitation and seniority on an equal and tunable footing provides a promising, flexible route for constructing systematically improvable CI expansions. With continuous weights, the approach can be extended by optimizing parameters for specific systems, incorporating orbital relaxation, or embedding in excited state and multilevel frameworks. This conceptual simplicity and adaptability make ehCI a compelling strategy for accurate and resource-efficient electronic structure calculations.

\begin{acknowledgement}
We acknowledge support from the National Science Foundation CAREER award CHE-2439867.
\end{acknowledgement}


\bibliography{achemso-demo}

\end{document}